\documentclass[final,3p,times,authoryear]{elsarticle}

\usepackage{amssymb}
\usepackage{graphicx}
\usepackage{subfig}
\usepackage{xcolor}
\usepackage{threeparttable}
\usepackage{blindtext}
\usepackage{booktabs}
\usepackage{makecell}
\usepackage{mathtools,amssymb}
\usepackage{amsmath}
\usepackage{multirow}
\usepackage{cases}
\usepackage{xfrac}
\usepackage{cite}
\usepackage{comment}
\usepackage{booktabs}
\usepackage{array}
\usepackage{natbib}
\usepackage{lscape}
\usepackage{adjustbox}
\usepackage{longtable}
\usepackage{tabularx}
\usepackage{float}
\DeclareGraphicsExtensions{.pdf,.png,.jpg}

\begin{document}
\begin{frontmatter}



\title{ Climate-Resilient Ports and Waterborne Transport Systems: Current Status and Future Prospects}

\author[label1]{Nadia Pourmohammad-Zia}
\ead{N.PourmohammadZia@tudelft.nl}
\affiliation[label1]{organization={Department of Hydraulic Engineering, Delft University of Technology, The Netherlands}}

\author[label1,label2]{Mark van Koningsveld}
\affiliation[label2]{organization={Van Oord Dredging and Marine Contractors, 3068 NH Rotterdam, The Netherlands}}

\begin{abstract}
The increasing challenges posed by climate change necessitate a comprehensive examination of the resilience of waterborne transport systems. This paper explores the critical nexus of climate resilience, waterborne transport, and technological innovation, addressing the challenges faced by ports and their connecting waterborne transport systems. It provides an in-depth analysis of the current status of climate-resilient infrastructure and operations while emphasizing the transformative potential of emerging technologies to improve resilience. Through a systematic review, the paper identifies critical gaps and opportunities. Research predominantly emphasizes port infrastructure over supply chain resilience, neglecting the interconnected vulnerabilities of maritime networks. There is limited focus on specific climate-induced disruptions, such as drought and compounded events, which complicate resilience planning. Methodologically, risk assessments and case studies dominate the field, while advanced technologies such as digital twins, artificial intelligence, and satellite monitoring remain underutilized. Geographic disparities in research output and a tendency toward short- to medium-term planning further constrain global and long-term resilience efforts. To address these gaps, the study advocates for systems-based approaches that integrate infrastructure, operations, and supply chains. It highlights collaborative frameworks and advanced tools, including digital twins, machine learning, and participatory modeling, as crucial for enabling predictive and adaptive risk management. This study stands as one of the first comprehensive reviews exclusively focused on climate resilience in ports and waterborne transport systems. It provides actionable insights for policymakers, researchers, and industry stakeholders, proposing a future research agenda to advance waterborne transport systems capable of withstanding multifaceted climate impacts.
\end{abstract}

\end{frontmatter}
\section{Introduction}
\label{intro}
\noindent From coastal terminals to inland waterways, waterborne transport systems form a dynamic circulatory network that underpins global trade. At the heart of this system are ports, which function as strategic gateways that concentrate flows of goods, people, and capital. These facilities, together with the inland waterborne connections that extend their reach into economic hinterlands, constitute a deeply interdependent infrastructure system. Their efficiency and reliability are not only vital for the smooth functioning of national economies but also for sustaining international supply chains \citep{saswat2024adoption}. As climate change progresses, these systems are increasingly exposed to a wide array of environmental stressors, including sea-level rise, drought, flooding, and extreme weather events. Understanding the resilience of ports and waterborne transport systems is therefore essential to maintaining the stability of global logistics and trade.\\
\indent Climate-related hazards introduce multifaceted risks to the continuity and performance of maritime and inland waterborne transport. Rising sea levels may reduce the viability of current port layouts and increase maintenance demands on critical assets. Changes in ocean dynamics and precipitation regimes can disrupt vessel schedules and reduce waterway navigability. For example, inland waterway transport, though energy-efficient, is particularly susceptible to drought, which can severely constrain vessel drafts and force operational delays. Recent events along the Rhine River have shown how localized hydrological disruptions can reverberate through continental supply chains. Additionally, fluctuations in cargo volumes linked to changes in agricultural production and shifting trade patterns add further complexity to long-term planning. The structural integrity of port infrastructure, such as quays, terminals, and storage facilities, may also be compromised by prolonged heat, saltwater intrusion, and variable precipitation, increasing the need for climate-aware asset management.\\
\indent In this context, resilience has become a central concept in the discourse on sustainable port and waterborne transport systems. It refers to the capacity of these systems not only to withstand shocks but also to adapt and recover effectively from them \citep{henry2012generic}. This includes both strategic measures that enhance long-term adaptability and operational mechanisms that enable rapid response during disruption. Building resilience requires a forward-looking approach to risk and an ability to manage both immediate hazards and evolving conditions over time.\\
\indent Disruptions affecting ports and waterborne transport systems stem from a variety of sources, including both environmental and anthropogenic events. These range from natural disasters such as floods and hurricanes to human-induced events like labor strikes and cyberattacks. While it is often useful to distinguish between types of disruptions for the purpose of planning, it is important to recognize that climate-related events can be either gradual or sudden. Drought, sea-level rise, and shifting seasonal patterns represent long-term pressures, whereas flash floods, wildfires, and heatwaves can emerge quickly with highly localized impacts. This variability calls for a resilience strategy that is both systemic and agile, combining robust infrastructure planning with flexible operational protocols.\\
\indent In parallel, recent technological advancements offer promising avenues for enhancing the resilience of maritime transport systems. Tools such as artificial intelligence, big data analytics, Internet of Things (IoT) devices, and digital twins are increasingly being explored for their potential to support climate risk management. These technologies may assist in anticipating disruptions, improving decision-making, and optimizing resource use. Examples include real-time data platforms for weather monitoring, predictive maintenance systems for infrastructure, and simulation models for stress-testing operational scenarios. At the same time, there is a growing need to examine methodological innovations that can enable better prediction, evaluation, and coordination of climate resilience strategies. However, the effectiveness of both technologies and methodologies is not yet fully understood and remains uneven across contexts. A more critical and integrative review is therefore needed to assess their actual contributions and identify where gaps remain.\\
\indent This paper addresses the following research question: How has existing research addressed the challenges of climate-induced disruptions in ports and waterborne transport systems, and what methodological, technological, and thematic gaps remain in advancing comprehensive, multi-scale, and forward-looking resilience strategies? To answer this, we review the full body of work on climate risks, resilience frameworks, and system-wide responses, with particular emphasis on two critical areas: the role of emerging technologies and the development of methodological approaches in resilience research. We examine how these tools are framed in the literature, what functions they are expected to serve, and the empirical evidence supporting their application. To enable a comprehensive response, we also clarify three foundational aspects: how resilience is defined and operationalized in this context; which key performance indicators (KPIs) are used to evaluate resilience, including system-level metrics and component-specific criteria; and which stakeholders are involved, along with the decision-making challenges they face. Our analysis synthesizes the current state of knowledge, highlights critical gaps, and proposes a future research agenda aimed at advancing intelligent, adaptive, and empirically grounded strategies for climate resilience in ports and waterborne transport systems. \\
\indent While previous literature reviews have explored related themes, our study builds upon and extends these efforts in distinct ways. \citet{gu2023systematic} provide a broad overview of maritime resilience but do not focus specifically on climate-induced disruptions or digital technologies. \citet{nguyen2023managing} examine maritime disruption management from an operational and organizational perspective, yet their review lacks a dedicated climate lens and is limited to firm-level analysis. In contrast, our review centers on climate resilience in both seaports and inland waterways and explicitly considers the dual role of advanced technologies and methodological tools in shaping more adaptive and scalable responses. This integrated perspective allows us to offer new insights into how climate resilience is framed in the literature and where opportunities exist for future research and application.\\
\indent The remainder of this paper is organized as follows: Section~\ref{term} defines resilience and explores its key characteristics. Section~\ref{RevMe} outlines the review methodology and presents a descriptive analysis. Section~\ref{Lit} reviews the state of the art, including climate-induced disruptions, technological advancements for climate resilience, key themes in the literature, and a comparative analysis of existing studies. Section~\ref{Dis} discusses current challenges and identifies opportunities for progress. Finally, Section~\ref{Conc} summarizes the findings and concludes the paper.
\section{Resilience: Definition and Characteristics}
\label{term}
\noindent Resilience has been conceptualized in a range of disciplines, each offering definitions tailored to the specific characteristics of the systems involved. In infrastructure research, resilience is often described as the capacity of a system to anticipate, absorb, adapt to, and recover from disruptive events \citep{resilience2010national}. In the context of supply chains, it has been defined as the adaptive ability to prepare for, respond to, and recover from unexpected disruptions while maintaining continuity and structural integrity \citep{ponomarov2009understanding}. Both definitions emphasize system-level functionality before, during, and after disruptions, and together offer a relevant foundation for understanding resilience in waterborne transport systems.\\
\indent In the literature, resilience is broken down into a number of related characteristics. However, terminology is not always used consistently across studies, and some terms are applied with different meanings or scopes depending on disciplinary background. In maritime transport, there remains limited clarity on how resilience-related concepts are defined and operationalized \citep{gu2023systematic}. To address this, Table~\ref{table-Term} summarizes commonly used resilience-related terms and their application in the context of infrastructure and transport systems.\\

\begin{table}[h]
    \caption{Description of Resilience-Related Terms}
    \label{table-Term}
    \renewcommand{\arraystretch}{1.4} 
    \centering
    \small
    \begin{tabular}{p{0.10\textwidth} p{0.33\textwidth} p{0.30\textwidth} p{0.17\textwidth}}
        \hline
        \textbf{Term} & \textbf{Description} & \textbf{Example} & \textbf{Reference} \\
        \hline
        Vulnerability & A characteristic that reduces or weakens the system’s ability to endure, manage, and survive disruptive events. & Low-lying terminals exposed to storm surges or sea-level rise. & \citet{liu2018analysis} \\ [1ex]
        Redundancy & The capacity of components to assume the functions of failed counterparts without affecting system performance. & Reserved berth capacity, backup cranes, or modal alternatives (e.g., shifting from barge to rail during drought). & \citet{kim2021framework} \\ [1ex]
        Adaptability & The ability to adjust configurations or operations in response to unforeseen changes. & Rerouting barges during low water levels or relocating operations to less exposed terminals. & \citet{leon2021adapting}, \citet{lavigne202317} \\ [1ex]
        Flexibility & The ability to reconfigure resources and cope with uncertainties through rapid adjustments. & Switching cargo to rail during flood-induced waterway closures. & \citet{leon2021adapting}, \citet{lavigne202317} \\ [1ex]
        Robustness & The capacity to resist disruption impacts and remain functional. & Quay walls engineered to withstand storm surges. & \citet{liu2018analysis} \\ [1ex]
        Reliability & The probability that a system maintains functionality despite disruptions. & Maintaining predictable barge schedules during mild drought due to effective water regulation. & \citet{asadabadi2020maritime} \\ [1ex]
        Recoverability & The ability to restore functionality quickly after disruption. & Clearing sedimentation from port access channels post-flood. & \citet{gu2023systematic} \\ [1ex]
        Rapidity & The speed of response and recovery. & Time to resume port operations after a hurricane. & \citet{salem2020probabilistic} \\ [1ex]
        \hline
    \end{tabular}
\end{table}

\begin{figure}[h]
\centering
\includegraphics[ clip, trim=3cm 14.8cm 1.5cm 3.45cm, width=0.75\textwidth]{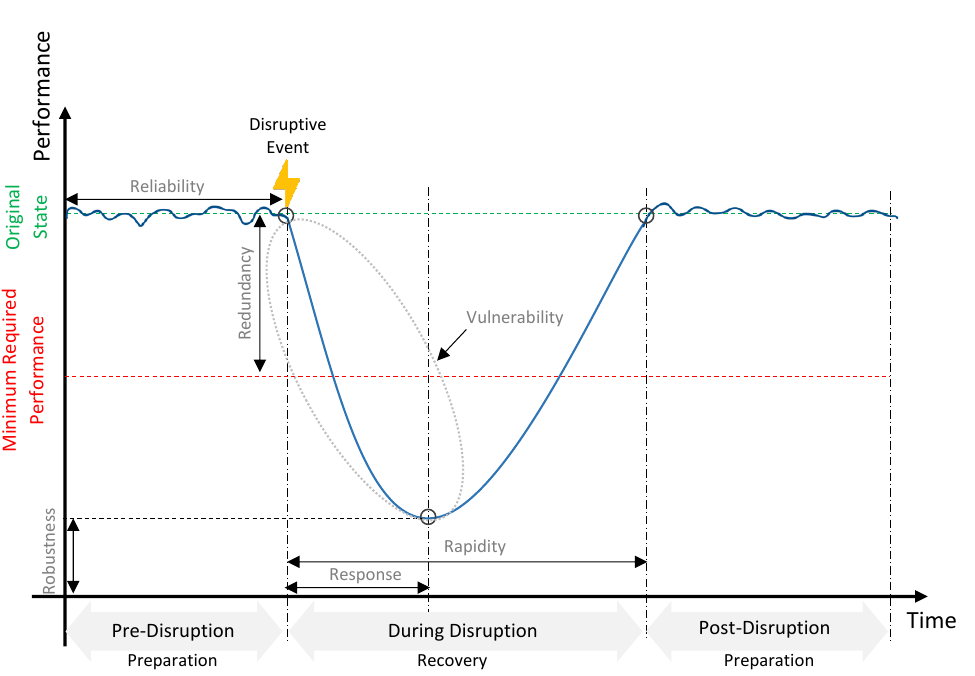}
\caption{Performance Diagram of a Resilient System (based on \citet{wan2018resilience})}
\label{Fig-ResDiag}
\end{figure}

\indent These characteristics also inform how we understand system performance over time. Figure~\ref{Fig-ResDiag} visualizes the performance trajectory of a resilient system before, during, and after a disruptive event.\\
\indent The performance of ports and waterborne transport systems is commonly assessed through four key indicators: capacity, efficiency, safety, and sustainability \citep{van2023ports}. These are reflected in tangible metrics such as throughput and berth availability (capacity), turnaround times and operational costs (efficiency), infrastructure reliability and incident prevention (safety), and environmental impact, including emissions and energy use (sustainability). Together, these measures offer a practical basis for evaluating system performance under both normal and disrupted conditions. During disruptions, these performance dimensions can deteriorate significantly before recovery processes restore functionality.\\ 
\indent The disruption timeline generally spans three phases: pre-disruption (stable operations), impact (performance degradation), and recovery. In the pre-disruption phase, systems are expected to operate near full capacity. This phase also represents the preparation stage, where proactive risk management and redundancy planning are essential. Once disruption occurs, system performance typically declines to a threshold level before stabilizing. The speed and extent of this decline depend on robustness, redundancy, and vulnerability. For example, vulnerability influences how quickly performance deteriorates, while robustness shapes how low performance may drop. During recovery, characteristics such as rapidity and recoverability determine how quickly and how fully performance is restored.\\
\indent Following recovery, the system may reach a new performance equilibrium. This equilibrium may differ from the original, either due to infrastructure upgrades or persistent impacts. For example, in the case of drought, even after water levels normalize, a loss of trust in inland waterway transport may persist, resulting in long-term shifts in modal choice or investment behavior. Thus, the full impact of disruption is not limited to physical conditions but includes psychological and behavioral effects as well.\\
 \citet{wan2018resilience} state that based on this analysis,  resilient ports and maritime transport systems should embody four primary characteristics, collectively referred to as 4R: reliability, redundancy, robustness, and recoverability. These attributes play a dominant role in determining the overall performance of the system, dictating its duration of steady performance, response to disruptive events, preservation of function post-disruption, and the attainment of a new equilibrium.
\section{Review Methodology and Descriptive Analysis}
\label{RevMe}
\noindent We employed a systematic procedure, searching for English-language publications published in peer-reviewed scientific journals from 2015 to 2025. Our search strategy involved querying the \textit{Scopus} and \textit{Web of Science} databases using a comprehensive Boolean search string, targeting terms that appeared in the title, abstract, or keywords fields. The search combined three thematic categories with logical \texttt{AND} operators:

\begin{enumerate} 
\item Resilience-related terms: \texttt{resilien* OR robust* OR adapt* OR reliabl* OR recover* OR\\ "climate-proof" OR "future-proof" OR flexibil* OR risk* OR vulnerabl*}; 
\item Transport-related terms: \texttt{port* OR terminal* OR seaport* OR "inland waterway*" OR\\ "inland shipping" OR "inland transport*" OR "waterborne transport*" OR harbor*\\ OR maritime OR "port-hinterland"}; 
\item Climate-related terms: \texttt{"climate change" OR climate OR flood* OR drought* OR "storm surge"\\ OR "sea level rise" OR "sea-level rise" OR "extreme weather" OR\\ "natural hazard*"}.  
\end{enumerate}

\begin{figure}[H]
\centering
    \includegraphics[ clip, trim=2.5cm 8.8cm 2.5cm 3.1cm, width=0.65\textwidth]{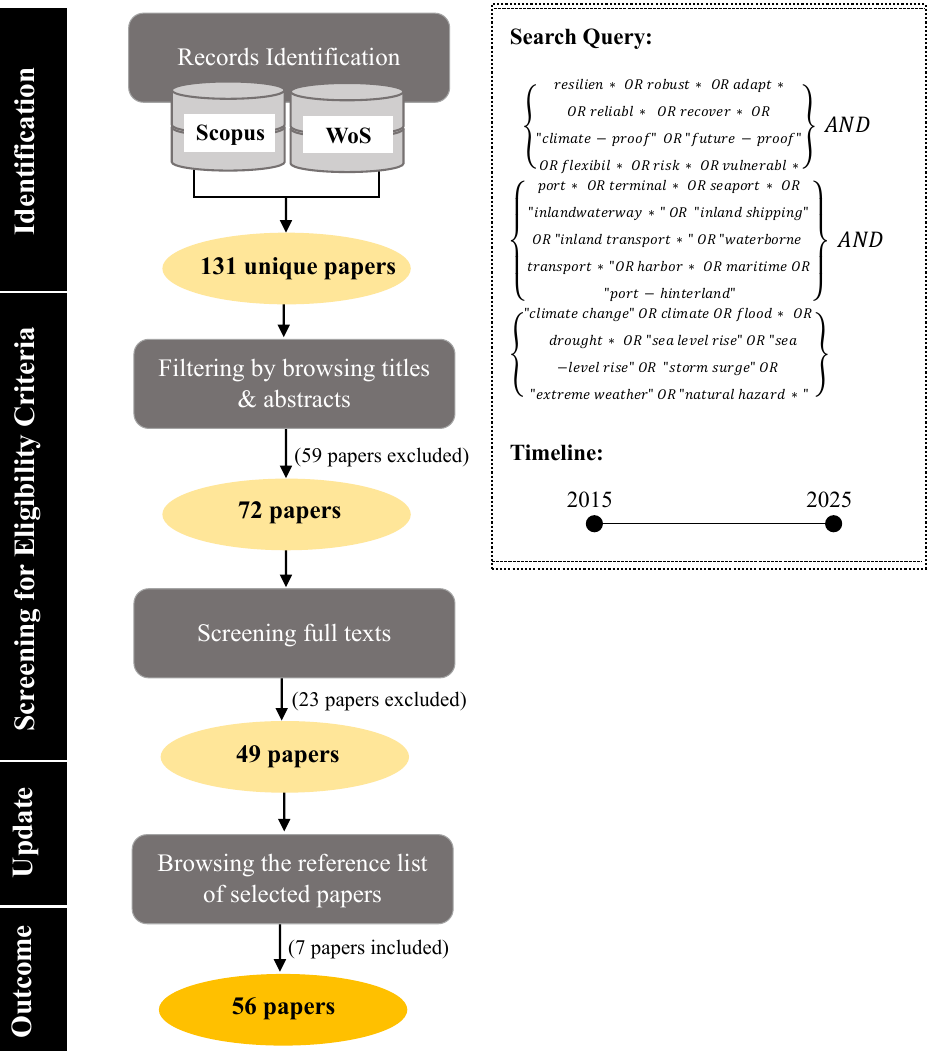}
    \caption{PRISMA flowchart for paper screening}
    \label{flowchart}
\end{figure}

\begin{figure}[H]
\centering
    \includegraphics[ clip, trim=0.2cm 1.5cm 0.3cm 0.55cm, width=0.75\textwidth]{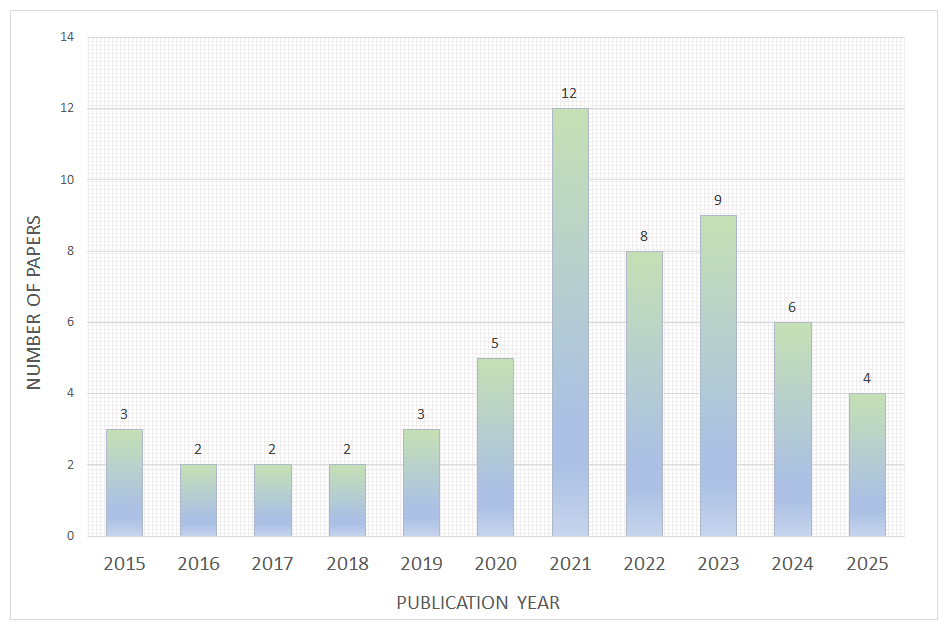}
    \caption{Distribution of The Papers by Their Publication Year}
    \label{Fig-Pubyear}
\end{figure}

\begin{figure}[h]
\centering
    \includegraphics[ clip, trim=0cm 14.1cm 0cm 0.1cm, width=0.7\textwidth]{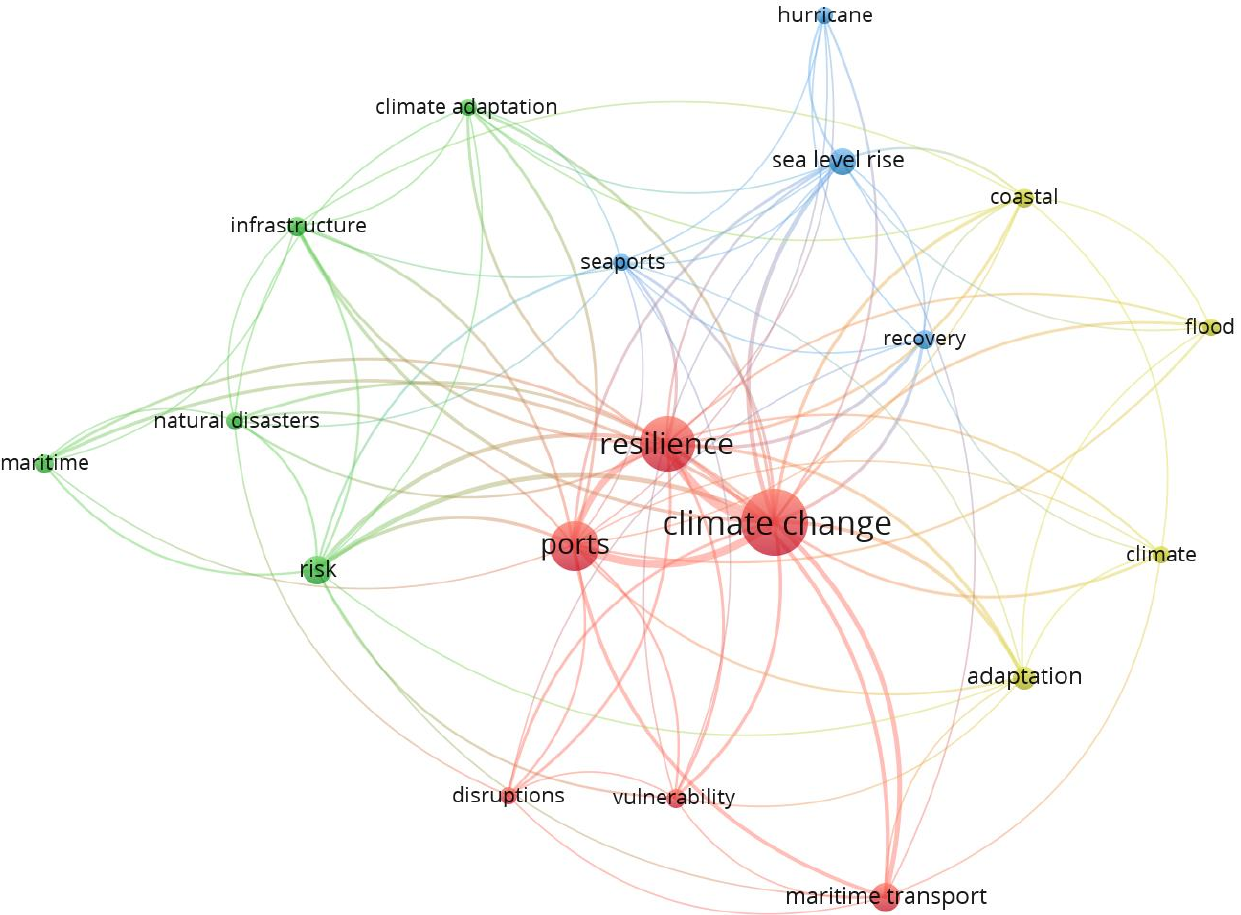}
    \caption{Co-Occurrence of Common Keywords in The Literature}
    \label{Fig-keywords}
\end{figure}

\begin{figure}[H]
\centering
    \includegraphics[ clip, trim=0.5cm 0.2cm 0.4cm 0.2cm, width=0.7\textwidth]{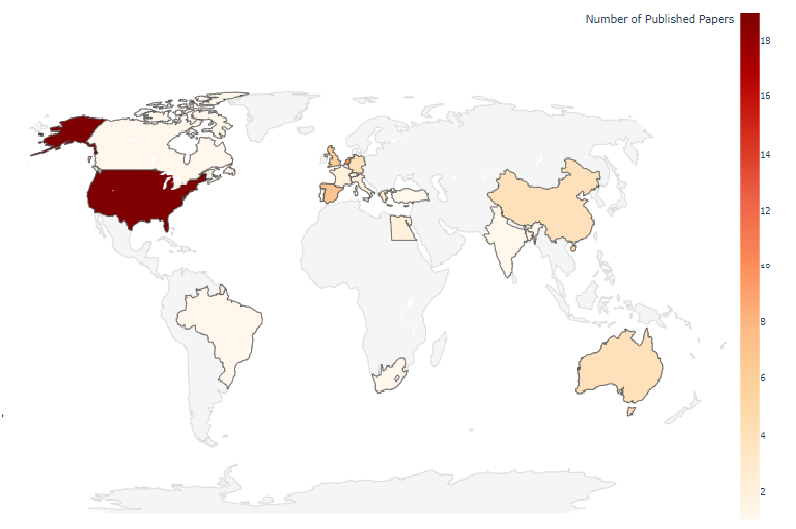}
    \caption{Research Output by Country}
    \label{Fig-Heatmap}
\end{figure}

\begin{figure}[H]
\centering
    \includegraphics[clip, trim=0.5cm 16.8cm 0.8cm 2.4cm, width=0.65\textwidth]{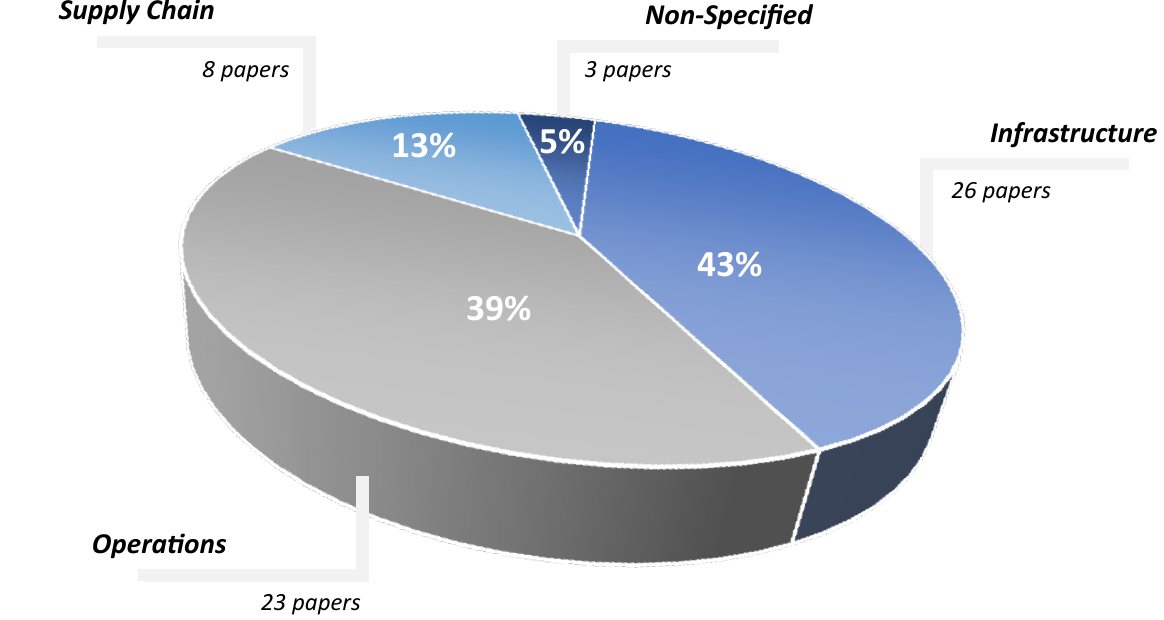}
    \caption{Share of Papers on Ports and Waterways Resilience Dimensions}
    \label{Fig-Dimensions}
\end{figure}

This structured approach ensured a focused and comprehensive review of literature relevant to climate-resilient port and waterborne transport systems. Figure \ref{flowchart} illustrates the PRISMA (Preferred Reporting Items for Systematic Reviews and Meta-Analyses) methodology adopted from \citet{moher2015preferred} that was employed for our paper screening.\\
\indent The initial search yielded 131 research papers, from which 59 irrelevant ones were excluded by screening the titles and abstracts. Exclusion criteria at this stage included: (1) papers that did not address port infrastructure, inland waterways, or maritime systems in the context of climate-related disruptions; (2) studies focused exclusively on land-based transportation (e.g., road or rail) with no connection to port, maritime logistics, or inland waterborne transport; (3) papers that, despite being indexed as journal articles, were in fact editorials, commentaries, or conceptual pieces lacking analytical depth or methodological rigor; (4) papers in which the term “port” referred to electronics, computer systems, or other unrelated technical contexts. Subsequently, we conducted a structured screening of the full text of the papers. This screening process yielded 49 publications that formed the basis for our in-depth analysis throughout the paper. We then investigated the reference list of these papers, leading to an addition of 7 papers to the basis for our further analysis.\\
\indent Building upon the outlined methodological foundation, we turn to a descriptive analysis of the selected literature to better understand the landscape of research on climate-resilient port and waterborne transport systems. This analysis provides a structured overview of the key characteristics of relevant studies, focusing on dimensions such as publication trends over time, keyword co-occurrence patterns, geographical distribution of research efforts, and the resilience dimensions addressed.\\
\indent Figure \ref{Fig-Pubyear}  represents the distribution of publication years across the literature surveyed, offering insights into the temporal trends and evolution of research in the field of climate-resilient port and waterborne transport systems.\\
\indent As shown in Figure \ref{Fig-Pubyear}, there is a gradual increase in publications from 2015 to 2020, followed by a significant surge in 2021. In subsequent years, the number of studies remains relatively steady, indicating sustained research interest in this area. The marked rise in 2021 may reflect a growing global awareness of the vulnerabilities posed by climate change to waterborne transport systems. This increase could also be linked to the intensification of climate-related disruptions during recent years, including extreme weather events, sea-level rise, and inland waterway droughts, which have highlighted the need for resilient solutions.\\ 
\indent Analyzing the keywords found throughout the literature provides an overview of common themes, patterns, and research priorities. To better understand these connections, we retrieved keywords from the reviewed papers and translated them into a co-occurrence diagram. Figure~\ref{Fig-keywords} illustrates the co-occurrence of keywords appearing in at least three articles, offering insights into the relationships and emphasis within the field.\\
\indent As depicted in Figure~\ref{Fig-keywords}, the size of each keyword node corresponds to its frequency of occurrence in the literature, with larger nodes indicating greater prominence. Keywords such as “Climate Change,” “Resilience,” and “Ports” emerge as the most dominant terms, visually shown by their larger size. “Maritime Transport” also appears prominently, reflecting its critical role in the broader research landscape. The connecting lines between keywords indicate co-occurrence, with the thickness of the lines representing the strength of these associations. “Climate Change,” “Resilience,” and “Ports” exhibit the strongest co-occurrence patterns, highlighting their centrality to the discourse on climate-resilient port and waterborne transport systems. In addition to general themes, the diagram highlights specific climate-related disruption types frequently addressed in the literature. “Sea Level Rise,” “Hurricane,” and “Flood” emerge as the most commonly studied disruptions, highlighting their immediate and severe impact on port operations and infrastructure. In contrast, terms such as “Drought” and “Wildfire” appear less frequently, indicating potential gaps in research. Similarly, resilience-related terminology such as “Risk,” “Vulnerability,” and “Recovery” is frequently mentioned, emphasizing the focus on understanding systemic weaknesses and developing adaptive strategies.\\
\indent Analyzing the geographic distribution of research output is essential for identifying regional disparities and highlighting areas requiring greater focus. Figure~\ref{Fig-Heatmap} depicts the global research output by country using a heatmap, which visualizes the number of papers published by researchers from each country.\\
\indent The heatmap in Figure~\ref{Fig-Heatmap} highlights a notable disparity in research output across countries, showing regional imbalances in academic focus on climate resilience. Some countries with large, strategically critical ports are among the top contributors to the field, while others, despite hosting major maritime hubs, exhibit relatively limited research activity. For instance, the United States, home to key ports such as the Port of Los Angeles and the Port of Long Beach, leads global contributions with 21 published papers, which reflects the country's prioritization of academic and policy-driven efforts to address the vulnerabilities of its waterborne transport systems to climate change. Following the United States, the Netherlands, which hosts the Port of Rotterdam, the largest port in Europe, has contributed 10 papers. This aligns with the Netherlands’ global reputation for leadership in water management and climate adaptation. Spain and the United Kingdom show notable research activities with 7 and 6 published papers, respectively. Germany, with the Port of Hamburg being one of Europe’s key ports, has a modest contribution of 4 papers. China, home to the largest port in the world, the Port of Shanghai, has contributed 4 papers. Furthermore, Singapore, which is the home to one of the world's busiest ports, the Port of Singapore, shows a relatively low research output with only 2 papers. This suggests a potential underutilization of its strategic and economic strengths in academic research on climate resilience. Interestingly absent from the higher tiers of research output in this dataset is South Korea, home to the Port of Busan, one of the largest ports globally. These findings highlight the need for increased collaboration and knowledge-sharing across regions, particularly in underrepresented countries that host critical ports.\\
\indent Ports and waterways resilience encompasses three dimensions: infrastructure, operations, and supply chain. When the infrastructure faces challenges, it hampers the smooth execution of operations. Accordingly, the vulnerability of the infrastructure and facilities directly impacts the operations and services.  Due to the complex nature of the maritime supply chain, disruptions in operations extend their impact beyond the port’s boundaries, affecting other components and the entire supply chain. Figure \ref{Fig-Dimensions} represents the share of papers focusing on each of these dimensions. \\
\indent As Figure \ref{Fig-Dimensions} illustrates, a significant portion, comprising 43\% of the papers, centers on infrastructure-related challenges. Following closely, 39\%  of the literature addresses operational aspects, emphasizing the key relationship between infrastructure and smooth operational execution. Additionally, 13\%  of the research focuses on the resilience of the whole supply chain, recognizing the broader impact of disruptions. When addressing the problem at the supply-chain level, fostering collaboration between ports emerges as a key factor in enhancing overall resilience through shared efforts and communication. This facet is acknowledged in only three papers within the existing literature.
\section{State of The Art}
\label{Lit}
\noindent This section synthesizes the existing body of academic research on climate resilience ports and waterborne transport systems. It is organized into thematic subsections that reflect key dimensions of the literature, including the types of climate-induced disruptions studied, the performance indicators used to assess resilience, the challenges faced by different stakeholders, the technological innovations supporting resilience, and the methodological approaches employed. 
\subsection{Climate-Induced Disruptions}
\noindent Climate change introduces a range of challenges for ports and waterborne transport systems, impacting their infrastructure and operations. As sea levels rise, low-lying port facilities face an increased risk of flooding, necessitating regular dredging to maintain navigational depths. The resultant sea level rise and changes in freshwater flows can lead to higher salinity levels in estuarine and coastal areas, affecting water quality and the structural integrity of port infrastructure, including piers and docks. The occurrence of intense storms, hurricanes, and typhoons poses a significant risk, causing physical damage to port facilities and temporarily suspending operations. Storm surges associated with these weather events contribute to coastal erosion and flooding, further emphasizing the vulnerability of port areas to extreme weather conditions. Extreme temperatures, including heatwaves, have practical implications for port operations, affecting the functionality of equipment and posing challenges to worker safety. Changes in precipitation patterns result in more extreme weather events, such as floods or droughts. Droughts, in particular, heighten the risk of low water levels, which can restrict vessel drafts. In response, vessels may be required to reduce their cargo loads to navigate through shallower waters, impacting the overall efficiency and cost-effectiveness of inland waterway transportation. Conversely, increased and intense rainfall leads to inland flooding, threatening infrastructure like locks, dams, and bridges, which are critical components of waterway systems. Alterations in ocean currents and wind patterns not only affect navigation routes but also increase the risk of accidents. Furthermore, the melting ice in polar regions may open up new shipping routes, presenting economic opportunities but also posing challenges to navigation safety and intensifying maritime activity in these previously inaccessible areas.\\
\indent Within the existing literature, a substantial portion, specifically 33 out of 56 papers, adopts a general perspective on climate-induced disruptions. These papers lack a focus on specific disruptions, encompassing a broader view of the challenges posed by climate change. While this approach provides a comprehensive understanding of the overall impact, it is essential to recognize that distinct disruptions often necessitate tailored strategies for enhancing resilience.\\
\indent Apart from the general perspective, a subset of papers focuses on specific climate-induced disruptions. Out of the 56 papers reviewed, 16 papers specifically address the impacts of sea level rise and flooding on port infrastructure and operations, which highlight the increased risk of global warming. Additionally, 2 papers focus on the effects of hurricanes, emphasizing the physical damage and operational disruptions caused by these intense storms. Interestingly, only five papers address the challenges posed by drought, which can lead to low water levels in inland waterways, restricting vessel drafts, and impacting the efficiency of transport. Similarly, two papers examine the implications of extreme wind events and storms, discussing how these conditions can affect navigation routes and increase the risk of accidents. This distribution of focus highlights the importance of broadening research efforts to cover a wider range of specific climate-induced disruptions, ensuring a more comprehensive understanding and preparedness for all potential impacts.
\subsection{Technological Advances in Service of Climate-Resilience}
\noindent \noindent As climate change poses unprecedented challenges to port and waterborne transport systems, innovative technologies can be seen as vital allies in the quest for resilience. These advancements are transforming how stakeholders approach climate adaptation, equipping them with the tools necessary to navigate the complexities of a changing environment.  Among the most promising developments are advanced predictive analytics and climate simulation models. These tools apply machine learning and big data to forecast climate impacts such as sea-level rise, storm surges, and extreme weather events, providing insights for proactive planning and decision-making. \\
\indent On the other hand, remote sensing and drones offer real-time monitoring capabilities, capturing detailed data on environmental conditions and infrastructure integrity. These technologies enhance our ability to assess and respond to climate-related challenges dynamically. Additionally, smart sensors and the Internet of Things (IoT) play a critical role by providing real-time data on factors such as temperature, humidity, and structural health, which helps in detecting and addressing potential issues promptly. Digital twinning is an emerging technology in this respect that incorporates smart sensors, IoT, artificial intelligence, etc., enabling real-time simulation and analysis of how climate change may affect infrastructure and operations. This allows for more accurate risk assessments and helps in planning effective adaptation measures.  Additionally, simulation and training tools, including virtual reality (VR) and augmented reality (AR), can be used to prepare personnel for climate-related scenarios and improve response strategies.\\
\indent The development of adaptive construction materials ensures that infrastructure can withstand extreme weather conditions, incorporating innovations such as corrosion-resistant materials and flexible, durable components. Furthermore, energy-efficient technologies, including renewable energy sources like solar and wind power, contribute to reducing the carbon footprint of port operations, aligning with sustainability goals while mitigating climate change impacts. Furthermore, automated and smart port systems enhance operational efficiency and resilience through technologies such as automated cargo handling and real-time traffic management.\\
\indent Despite significant technological advancements and the widespread integration of these technologies in weather forecasting, their application in enhancing climate resilience for ports and inland waterways remains in its early stages. The literature is relatively sparse, with few papers exploring the integration of these technologies into climate adaptation and resilience strategies. Among these few papers are the studies conducted by \citet{tsaimou2023uav} and \citet{zhou2021analytics}, which investigate the capabilities of an Unmanned Aerial Vehicle (UAV) monitoring program and a Decision Support System with digital twinning-based resilience analysis, respectively. Another interesting case is the open-source Digital Twin of the Dutch waterways developed by \citet{Deltares2023}. This digital twin integrates comprehensive data on infrastructure, hydrodynamics, and vessel behavior to simulate and assess the potential effects of climate change and the corresponding adaptation strategies. By modeling both infrastructural and logistical interventions, it provides insights into how these factors influence the performance and resilience of inland waterways.\\
\indent Four recently launched EU projects, \citet{Clarion}, \citet{Safari}, \citet{Cristal}, and \citet{ReNEW}, are poised to make significant contributions to the emerging body of research aimed at strengthening the resilience of port and inland waterway systems through innovative technologies. The \textit{CLARION} project focuses on integrating UAVs, satellite-based and sensor-driven environmental measurements, smart and sustainable quay walls, and digital twins for flood impact management and optimization of multimodal port-hinterland connections. Similarly, the \textit{SAFARI Ports} project implements advanced monitoring and predictive technologies, including digital twinning, to enhance resilience and prevent structural failures in the ports of Seville, Dunkirk, and Lisbon. The \textit{CRISTAL} project aims to boost inland waterway transport freight share by at least 20\% and maintain 50\% system capacity even under extreme weather conditions. It demonstrates integrated technological and governance solutions at pilot sites in Italy, France, and Poland, focusing on digitalization, infrastructure resilience, and alignment with Physical Internet principles. The \textit{ReNEW} project complements this by targeting climate-neutral and climate-resilient IWT through innovations in vessel fleet reliability, infrastructure sustainability, digital integration, and workforce development, aiming to validate how digital and energy transitions can support a high-efficiency, low-emission inland waterway system.
\subsection{Key Methodological Themes}
\begin{figure}[h]
\centering
    \includegraphics[ clip, trim=0.1cm 0.2cm 0.1cm 0.5cm, width=0.8\textwidth]{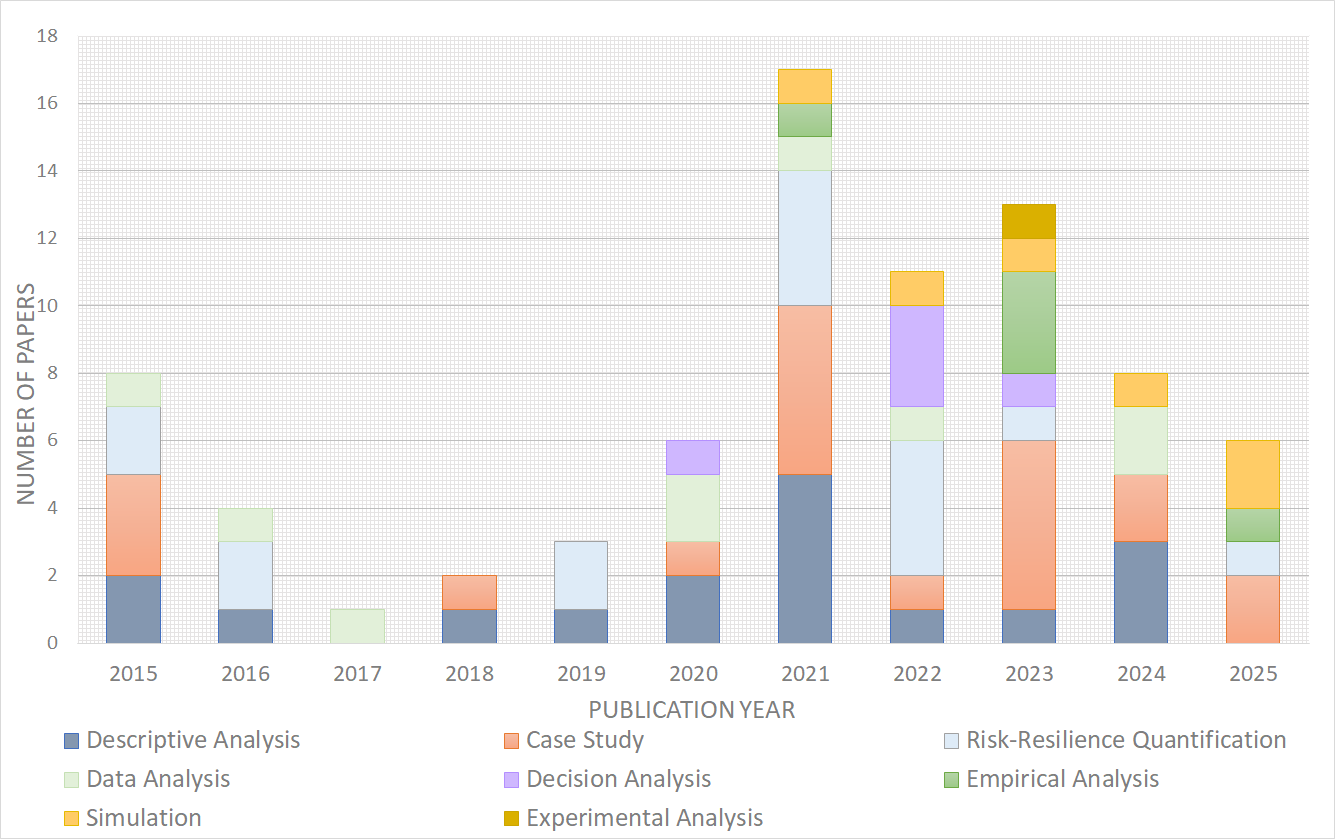}
    \caption{Share of Applied Methodologies}
    \label{Fig-Methods}
\end{figure}

\indent A diverse range of methodological approaches has been incorporated into the study of climate-induced disruptions in the port-hinterland landscape. These methods include Descriptive Analysis, which explores concepts and theories to offer comprehensive insights; Case Studies, which investigate specific instances to reveal practical implications; Risk-Resilience Quantification, which measures the resilience of port systems using various indices; Data Analysis, employing Statistical methods, Time Series Analysis, and Machine Learning to extract insights from datasets; Decision Analysis, utilizing tools like Game Theory, Mathematical Programming, and Multi-Criteria Decision Analysis (MCDA) to inform strategic choices; Empirical Analysis, focusing on validation through Surveys, Questionnaires, Interviews, Focus Groups, and Expert Consultations; Simulation, modeling complex scenarios and interactions; and finally, Experimental Analysis, using controlled experiments to validate hypotheses. Figure \ref{Fig-Methods} illustrates the distribution of these methodologies across various years from 2015 to 2025, providing insights into evolving trends, methodological strengths, and potential research gaps over the decade. In some cases, papers integrate multiple approaches to offer a more comprehensive analysis.\\
\indent Analyzing the methods applied reveals that Descriptive Analysis and Case Studies are the most frequently used approaches, with significant spikes in 2021. This indicates a strong emphasis on qualitative insights and detailed examinations of specific instances during that year. Risk-resilience Quantification has seen an increasing trend in 2021 and 2022. Data analysis is consistently utilized across the years, suggesting a steady reliance on quantitative approaches to understand and predict climate impacts. However, Decision Analysis appears less frequently, with mentions only in 2020, 2022, and 2023, highlighting a potential gap in strategic decision-making studies related to enhancing climate resilience. Empirical and Experimental Analysis methods are rarely employed, pointing to a need for more research to validate theoretical models. Finally, Simulation-Based methods made their first appearance in 2021 and have been used regularly since, which indicates a growing recognition of their value in modeling complex climate scenarios. The diversification of methodologies in recent years, especially from 2020 onwards, marks a maturing field increasingly embracing quantitative and analytical approaches. However, the initial years were dominated by descriptive and case study approaches, indicating an early focus on exploring and understanding phenomena rather than quantifying or modeling them.\\
\indent A closer look at the literature reveals that descriptive analysis and case studies are often employed alongside each other, making them the most prevalent approach for examining climate resilience within port and maritime systems. This pairing highlights the preference for qualitative approaches, where researchers often combine detailed descriptions with in-depth examinations. As fairly early research works, \citet{dircke2015climate} examines strategic adaptation measures in Rotterdam, while \citet{becker2015stakeholder} focus on seaport resilience through a Gulfport case study. Both studies highlight specific geographical contexts to provide insights into resilience strategies. Similarly, \citet{trundle2019leveraging} and \citet{wood2019climate} analyze climate impacts using case studies from Pacific islands and Hurricane Sandy, respectively, emphasizing the importance of local perspectives and disaster implications on urban resilience.\\
\indent This focus on localized disruptions extends to more recent research, which analyzes the effects of climate-induced disruptions on port operations and infrastructure. For example, \citet{ryan2020port} and \citet{ribeiro2021flooding} study the impacts of Hurricane Sandy and flood risks at Aveiro Port, respectively. These studies offer detailed assessments of the mentioned disruptive events on the operations of Aveiro Port. \citet{jarriel2021climate} extends this analysis by examining ancient maritime networks' adaptation strategies in the Bronze Age Cyclades, while \citet{echendu2021relationship} explores the link between urban planning and flooding in Port Harcourt. \citet{abdelhafez2021vulnerability} review tropical cyclones' effects on Gulf Coast seaports, and \citet{da2022climate} assess how Brazilian ports integrate climate change into management policies. \citet{zittis2023maritime} use advanced climate data to evaluate future maritime risks in European islands, and \citet{sharaan2024qualitative} investigate adaptation strategies employed by Egyptian port authorities. Lastly, \citet{toledo2024nature} propose a nature-based solution with vegetated dunes to mitigate flooding in Benidorm.\\
\indent In addition to combination with case studies, descriptive analysis has also been employed independently across a broader range of research areas. In that respect, \citet{morris2020stakeholder} provides a comprehensive overview of port climate adaptation approaches, emphasizing the importance of collaboration among coastal stakeholders in implementing climate adaptation strategies within port communities. The study highlights the lack of federal support for climate adaptation in the maritime sector and suggests that novel methods are needed to facilitate adaptation efforts at both individual port and regional levels. Similarly, \citet{thakur2021ports} explores the conceptual framework surrounding seaports, focusing on the challenges posed by climate change. The research goes over theoretical debates on climate uncertainty, policymaking, and path dependencies in port planning. It also discusses global best practices, knowledge gaps, and strategies that ports can adopt to better confront climate-induced disruptions in the future. Lastly, \citet{brooke2024port} emphasize the utility of adaptation pathways, which allow for phased responses to evolving climate hazards such as sea-level rise and flooding. By identifying a range of future climate projections, this approach aids in developing more flexible and responsive adaptation strategies for port communities.\\
\indent Another significant theme in the literature is the quantitative assessment of risk and resilience. Various frameworks and indices have been developed to quantitatively understand and enhance resilience. \citet{neumann2015future} quantify the impacts of climate-induced disruptions on coastal populations, while \citet{testa2015resilience} estimate the resilience of coastal transportation networks in New York. \citet{lam2016quality} develop a quality function deployment approach to enhance maritime supply chain resilience, and \citet{hosseini2016modeling} quantify resilience using Bayesian networks to evaluate absorptive, adaptive, and restorative capacities. \citet{zhou2021analytics} introduce a Decision Support System with digital twinning-based resilience analysis. Specific indices have also been developed to measure port resilience, including \citet{kontogianni2019development}, \citet{leon2021adapting}, \citet{poo2021climate}, \citet{verschuur2022systemic}, and \citet{liu2023analysis}. \\ 
\indent Other studies extend this quantitative focus by evaluating and quantifying disruption risks. \citet{wan2022evaluating} establish a risk-based framework for maritime logistics and supply networks, and \citet{poo2024optimising} combine climate risk indicators with ship routing optimization to assess global shipping network resilience. Moreover, \citet{christodoulou2019sea} analyze risks associated with sea-level rise projections for seaports.  \citet{izaguirre2021climate} apply the IPCC’s risk-modeling framework to assess future port operation risks under high-end emissions pathways. Lastly, \citet{balakrishnan2022methodology} propose a methodology to analyze economic risks of hurricane-related port shutdowns, using regression analysis and input-output models.\\ 
\indent As the third common theme in the literature, data analysis is essential for quantifying and predicting the impacts of climate change on port infrastructure and operations. By utilizing diverse datasets, these studies offer valuable insights into patterns, trends, and emerging risks, enabling the development of targeted and data-driven adaptation strategies. As early research in this category, \citet{becker2016method} estimate the demand for construction materials needed to protect seaports from sea-level-rise-enhanced storm surges. Their study calculated resource requirements for a coastal storm surge protection structure suited to upper-bound projections of two meters of sea level rise by 2100, demonstrating how precise data can guide infrastructure planning. Similarly, \citet{allen2022sea} develop a methodology to assess the exposure of major container port terminals in the U.S. to sea level rise. By incorporating geospatial data, elevation models, tide gauges, and sea level projections, the study enables a comparative assessment of risks to port infrastructure and associated surface transportation, providing a clear example of how diverse datasets can inform vulnerability assessments. In line with these approaches, \citet{grant2024no} employ network analysis to explore Port Miami’s regional maritime connections. Using multiple years of marine traffic and cargo data, the study identified key maritime partners and assessed their potential impact on the port’s resilience.\\
\indent Further contributing to the theme, \citet{van2020effect} presented a model to assess the effects of reduced water depth on inland ships' deadweight capacity, utilizing a detailed table of minimum required drafts and under keel clearance for different ship types and sizes. This type of analysis, similar to \citet{becker2016method}, underscores the role of quantitative data in addressing specific operational challenges linked to climate impacts. Other papers use various data sources to analyze past and potential disruptions. \citet{verschuur2020port} employed vessel tracking data to examine past port disruptions caused by natural disasters, illustrating how big data can provide insights into operational vulnerabilities. In contrast, \citet{moreno2021resilience} evaluated historical management strategies through the collection of human and natural system data, offering a retrospective analysis that informs future resilience efforts. Lastly, \citet{vinke2024inland} analyzed ten years of IVS and discharge data to explore the relationship between discharge levels and vessel deployment, offering a new level of detail in understanding inland waterway challenges in the face of drought. \\
\indent Decision analysis significantly influences strategic decision-making for enhancing resilience against climate disruptions. By offering a structured approach to evaluate and compare adaptation strategies, it helps stakeholders manage trade-offs and prioritize investments. For instance, \citet{asadabadi2020maritime} develop a stochastic, bi-level game-theoretic optimization model to assess and improve the resilience and reliability of global port networks. This model, grounded in decision theory, helps stakeholders navigate complex global interactions and optimize strategies to bolster port resilience amidst uncertainties. \citet{li2022enhancing} apply a bi-level equilibrium framework, proposing strategies such as capacity sharing and cross-port investments to enhance the resilience of individual ports and the overall network. Expanding on this collaborative decision framework, \citet{shi2023improving} examine the strategic dynamics of competing maritime supply chains, analyzing the impact of cooperation between ports, inland logistics providers, and government regulation. The study highlights the importance of cooperation and regulatory influence in enhancing supply chain resilience. Additionally, \citet{rubinstein2023nature} introduce an innovative decision-making methodology using the MoSCoW framework and multi-criteria analysis (MCA) to assess the feasibility and site selection for nature-based solutions (NBM) in ports. This approach, similar to the multi-criteria considerations in \citet{asadabadi2020maritime}, helps stakeholders systematically prioritize projects based on essential criteria, balancing short-term needs with long-term resilience goals. \\
\indent Next is empirical analysis, which plays an important role in validating theoretical models and deepening our understanding of climate resilience in port and hinterland systems through direct engagement with stakeholders and experts. Utilizing methods such as surveys, questionnaires, interviews, focus groups, and expert consultations, empirical research not only tests theoretical assumptions but also captures the nuanced experiences and perspectives of those directly involved in or impacted by port and hinterland operations and climate adaptation efforts. One notable study, \citet{mclean2021advancing}, utilize semi-structured interviews to explore seaport decision-makers' perceptions of strategies for overcoming adaptation barriers. This research offers valuable insights into the practical challenges and effectiveness of different adaptation approaches from the perspective of those directly involved in decision-making processes. Similarly, \citet{islam2023developing} engage with a marine stakeholder and a marine geologist to create and evaluate a catastrophe scenario, demonstrating the role of expert consultations in assessing the real-world feasibility of theoretical models. In another example, \citet{chowdhooree2024climate} combine focus group discussions, GIS-based mapping, and key informant interviews to understand community responses to climate changes. This comprehensive approach provides a detailed view of how spatial development and community perceptions intersect with climate adaptation. Additionally, \citet{fouad2023waterways} conduct surveys to assess public perceptions of the transformation of Alexandria's historic canal, emphasizing the importance of community awareness and engagement in addressing vulnerabilities and adapting to climate change.\\
\indent Simulation is instrumental in examining complex climate scenarios and their impacts on port and hinterland systems. By creating virtual models of port operations, infrastructure, and environmental conditions, simulation enables researchers to explore how various factors, such as extreme weather events, sea-level rise, or shifts in trade patterns, affect the resilience and efficiency of these systems. One study, \citet{bal2023towards}, address the potential role of railways in freight transport if the Rhine, a major European waterway, faced local inaccessibility due to extremely low water levels. This research, prompted by the 2022 dry summer, utilized simulation to assess whether rail could temporarily replace river transport. Another investigation, \citet{folkman2021port}, employ Arena to simulate port operations before and after a hurricane, focusing on the impact of de-ballasting on key performance indicators. The simulation sheds light on how extreme weather events can affect port performance and informs strategies for operational adjustments. In addition, \citet{vinke2022cascading} introduce a simulation method to incorporate the cascading effects of low discharge events in climate risk assessments for waterborne supply chains. The study demonstrated how these effects and corresponding mitigation measures can be modeled to understand their impact on supply chain performance. Finally, \citet{Bakker2024PAINC} combine a mesoscale logistical model with a semi-empirical hydrodynamic model to analyze lock operations and saltwater intrusion in the face of droughts. Their approach highlights how simulation can integrate operational logistics with environmental changes to provide a detailed view of their interactions.\\
\indent Experimental analysis provides a hands-on approach to understanding climate resilience by directly testing theories and models in controlled or real-world environments. Through experiments, researchers can validate hypotheses, refine models, and evaluate the effectiveness of adaptation strategies in a tangible setting. However, due to its inherent complexity and high costs, experimental analysis is less prevalent in the literature, with only one study currently utilizing this approach. The study by \citet{tsaimou2023uav} exemplifies the use of experimental analysis in advancing our understanding of port infrastructure resilience. The research focuses on investigating the capabilities of an Unmanned Aerial Vehicle (UAV) monitoring program to enhance Structural Health Monitoring practices for port infrastructure. The study involved four in-situ inspections conducted by the Laboratory of Harbour Works of the National Technical University of Athens at Lavrio Port in central-eastern Greece. 
\subsection{Comparative Overview of the Literature}
\noindent Table \ref{tab:literature_review} offers a comprehensive overview of existing research on climate resilience in port infrastructure, operations, and supply chains. The table systematically categorizes papers based on key headings: Research Focus, Scope, Dimension, Disruption Type, Scale, Horizon, and applied methodologies. ``Research Focus'' highlights the main objective of the study, such as risk assessment, adaptation planning, or policy review. ``Scope'' identifies the domain under investigation, such as ports, inland waterways (IWs), or the maritime supply chain (SC). ``Dimension'' indicates whether the study focuses on infrastructure, operations, or the whole supply chain. ``Disruption Type'' specifies the climate-induced challenges addressed, such as sea-level rise, flooding, hurricanes, etc. ``Scale'' defines the geographical context, whether at local, regional, national, or global levels. ``Horizon'' describes the time frame considered, such as short-term, medium-term, or long-term impacts. Finally, ``Methodology'' outlines the research approach applied.\\
\indent This structured approach provides a bird’s-eye view of trends, strengths, and critical gaps within the current body of literature. It enables a clearer understanding of where research efforts have been concentrated, the methods employed, and where significant opportunities for future investigation still exist.
\begin{landscape} 
\begin{longtable}[c]{>{\raggedright\arraybackslash}p{2.8cm} >{\raggedright\arraybackslash}p{2.9cm} >{\raggedright\arraybackslash}p{1.9cm} c>{\raggedright\arraybackslash}p{2cm} c >{\raggedright\arraybackslash}p{2cm} >{\raggedright\arraybackslash}p{4.1cm}}
    \caption{Comparative Literature Review Table} \label{tab:literature_review} \\
    \toprule
    \textbf{Reference} & \textbf{Research Focus} & \textbf{Scope} & \textbf{Dimension} & \textbf{Disruption} & \textbf{Scale} & \textbf{Horizon} & \textbf{Methodology} \\
    \midrule
    \endfirsthead 

   \caption*{Comparative Literature Review Table (continued)} \\ 
    \toprule
    \textbf{Reference} & \textbf{Research Focus} & \textbf{Scope} & \textbf{Dimension} & \textbf{Disruption} & \textbf{Scale} & \textbf{Horizon} & \textbf{Methodology} \\
    \midrule
    \endhead 

    \midrule
    \multicolumn{8}{r}{\textit{Continued on next page}} \\ 
    \endfoot 

    \bottomrule
    \endlastfoot 

    \citet{becker2015stakeholder} & Policy Review & Ports & - & General & National & Long-term & Stakeholder Survey\\   
    \citet{dircke2015climate} & Policy Review & Ports & Infra & Flooding & Local & Medium-term & Case Study \\     
    \citet{neumann2015future} & Vulnerability Assmt & Ports & Infra & Sea level Rise \& Flooding & Global & Long-term & Scenario-based Exposure Modelling \\    
    \citet{testa2015resilience} & Hazard Mitig. & Port's Net. & Infra & General & Regional & Short \& Medium-term & Topological Graph-based Network Analysis \\     
    \citet{becker2016method} & Adaptation Plan. & Ports & Infra & Sea level Rise & Global & Long-term & Forecasting \& Resource Estimation \\     
    \citet{hosseini2016modeling} & Resilience Assmt & IWs & Infra & General & - & Short \& Medium-term & Bayesian Networks Modelling \\ 
    \citet{lam2016quality} & Policy Review & Maritime SC & SC & General & - & - & Empirical Data Analysis \\ 
    \citet{repetto2017integrated} & Risk Assmt & Ports & Oper & General & Regional & Short \& Long-term & Monitoring Network \& Forecasting \\    
    \citet{christodoulou2019sea} & Risk Assmt & Ports & Oper & Extreme Win Events & Global & Long-term & Comparative Analysis \\     
    \citet{kontogianni2019development} & Vulnerability Assmt & Ports & Infra & Sea level Rise \& Flooding & Local & Short \& Medium-term & Empirical Data Analysis \\   
    \citet{trundle2019leveraging} & Vulnerability Assmt & Ports & Infra & General & Regional & Short \& Medium-term & Community Consultations \& Stakeholder Survey \\ 
    \citet{wood2019climate} & Policy Review & Ports & - & Sea level Rise & Local & Long-term & Theoretical Analysis \\    
    \citet{asadabadi2020maritime} & Risk Assmt & Port's Net. & SC & General & Global & Long-term & Stochastic, Game-Theoretic Optimization \\ 
    \citet{morris2020stakeholder} & Policy Review & Ports & Infra/Oper & General & Regional & Long-term & Literature Review \& Case Study \\   
    \citet{ryan2020port} & Risk Assmt & Ports & Infra/Oper & Extreme Wind Events & Local & Short-term & Survey \& Case Study Analysis \\    
    \citet{van2020effect} & Adaptation Plan. & IWs & Oper & Drought & Regional & Medium \& Long-term & Field Observations \& Empirical Data Analysis \\ 
    \citet{verschuur2020port} & Risk Assmt & Ports & Oper & General & Regional & Long-term & Empirical Data Analysis \\  
    \citet{abdelhafez2021vulnerability} & Risk Assmt & Ports & Infra/Oper & Sea level Rise \& Hurricanes & Regional & Long-term & Fault Tree Analysis \\    
    \citet{echendu2021relationship} & Hazard Mitig. & Ports & Infra & Flooding & Local & Short-term & Stakeholder Survey \\    
    \citet{folkman2021port} & Risk Assmt & Ports & Oper & Hurricanes & Local & Short-term & Simulation Analysis \\   
    \citet{izaguirre2021climate} & Risk Assmt & Ports & Oper & General & Global & Long-term & IPCC Risk Model \\  
    \citet{jarriel2021climate} & Risk Assmt & Port's Net. & Infra & General & Local & Long-term & GIS \& Archaeological Data Analysis \\     
    \citet{leon2021adapting} & Resilience Assmt & Port's Net. & SC & General & Local & Long-term & Resilience Index \& Case Study Analysis \\  
    \citet{mclean2021advancing} & Adaptation Plan. & Ports & Oper & General & Regional & Medium-term & Semi-Structured Interviews \\      
    \citet{poo2021climate} & Risk Assmt & Ports & Infra/Oper & General & National & Short \& Long-term & Quantitative Risk Analysis (Evidence Reasoning) \\    
    \citet{ribeiro2021flooding} & Risk Assmt & Ports & Infra & Flooding & Local & Long-term & Spatial Mapping \& Simulation Analysis \\  
    \citet{thakur2021ports} & Policy Review & Ports & Infra/Oper & General & National & Long-term & Case Study Analysis \\   
    \citet{zhou2021analytics} & Resilience Assmt & Ports & Infra & General & Local & Medium-term & Digital Twinning \\    
    \citet{allen2022sea} & Adaptation Plan. & Ports & Infra/Oper & Sea level Rise & Regional & Long-term & Geospatial Data Analysis \\    
    \citet{balakrishnan2022methodology} & Risk Assmt & Ports & Infra & Hurricanes & Regional & Short \& Medium-term & Regression Analysis \& Input-Output Modeling \\
    \citet{da2022climate} & Policy Review & Ports & Infra & General & Multi-Scale & Short-term & Literature Review \& Documents Analysis \\
    \citet{li2022enhancing} & Adaptation Plan. & Net. of Ports & SC & General & Regional & Long-term & Game Theory \\
    \citet{verschuur2022systemic} & Risk Assmt & Maritime SC & SC & General & Multi-Scale & Long-term & Risk Framework Development \\
    \citet{vinke2022cascading} &  Hazard Mitig. & IWs & SC & Drought & Regional & Long-term & Data Analysis \& Simulation \\
    \citet{wan2022evaluating} & Adaptation Plan. & Port's Net. & Oper & General & Global & Short \& Medium-term & Comparative \& Resilience Loss Triangle Analysis \\
    \citet{bal2023towards} & Recovery Plan. & Port's Net. & Oper & Drought & Regional & Short-term & Simulation \& Case Study Analysis \\
    \citet{fouad2023waterways} & Policy Review & Ports & Infra & General & Local & Short \& Medium-term & Strategy Development \\
    \citet{islam2023developing} & Recovery Plan. & Port's Net. & Oper & General & Regional & Short \& Medium-term & Stakeholder Survey \& Scenario Development \\
    \citet{liu2023analysis} & Resilience Assmt & Net. of Ports & Oper & General & International & Medium \& Long-term & Graph Theory, Topological Indices \& Simulation \\
    \citet{rubinstein2023nature} & Adaptation Plan. & Ports & Infra & Flooding & Regional & Short \& Medium-term & Multi-Criteria Analysis \\
    \citet{shi2023improving} & Resilience Assmt & Net. of Ports & SC & General & National & Short \& Medium-term & Game Theory \\
    \citet{tsaimou2023uav} & Adaptation Plan. & Ports & Infra & General & Local & Short \& Medium-term & Empirical \& Data Analysis \\
    \citet{zittis2023maritime} & Risk Assmt & Ports & Infra/Oper & General & International & Long-term & Data Analysis \& Impact Chain Approach \\
    \citet{Bakker2024PAINC} & Risk Assmt & Ports & Infra & Drought & Local & Short \& Medium-term & Hydrodynamic Modeling \& Simulation \\
    \citet{brooke2024port} & Adaptation Plan. & Ports \& IWs & Infra/Oper & General & - & Long-term & Scenario-Based Risk Assessment \& Data Analysis \\
    \citet{chowdhooree2024climate} & Policy Review & Ports & Infra & Drought & Local & Short-term & Survey \& Case Study Analysis \\
    \citet{grant2024no} & Resilience Assmt & Net. of Ports & SC & General & Regional & Medium \& Long-term & Data-Driven Network Analysis \\
    \citet{poo2024optimising} & Adaptation Plan. & Port's Net. & Oper & General & Global & Long-term & Risk Indicators \& Shipping Route Optimization \\
    \citet{sharaan2024qualitative} & Risk Assmt & Ports & Infra & General & National & Medium-term & Survey \& Group Discussions \\
    \citet{toledo2024nature} & Adaptation Plan. & Ports & Infra & Flooding & Local & Short \& Medium-term & Case Study Analysis \\
    \citet{vinke2024inland} & Risk Assmt & IWs & Oper & Drought & Regional & Medium-term & AIS \& IVS Data Analysis \\   
    \citet{romya2025adaptation} & Adaptation Plan. & Ports & infra & Sea level Rise & National & Medium-term & Hydrodynamic Modelling \\
    \citet{polydoropoulou2025living} & Policy Review & Ports & Infra/Oper & General & Local & Medium \& Long-term & Survey \& Case Study \\
    \citet{johnson2025flood} & Risk Assmt & IWs & infra & Flooding & National & Medium \& Long-term & Agent-based \& Economic Input-Output Modelling \\
   \citet{ahmadi2025current} & Risk Assmt & IWs & Infra/Oper & Flooding & National & Medium \& Long-term & Strategic Resilience Index \\    
\end{longtable}
\end{landscape}
\noindent The analysis of the table reveals several important trends in existing research. The majority of studies focus on port infrastructure as the primary dimension, emphasizing the vulnerability of physical assets to climate-induced disruptions. These disruptions include frequently investigated phenomena such as sea-level rise, flooding, and extreme weather events. However, the relatively narrow focus on infrastructure often neglects the interconnected nature of port systems, which rely heavily on operational efficiency and supply chain continuity. Studies addressing drought remain underrepresented, highlighting a critical gap in understanding how water scarcity affects inland waterways, port operations, and the broader maritime logistics network. This underrepresentation is particularly concerning given the growing prevalence of extreme drought events globally, such as those that have recently disrupted major inland waterway systems like the Rhine and Mississippi Rivers.\\
\indent Additionally, most research is concentrated at local and regional scales, often addressing site-specific vulnerabilities and solutions. While such studies are valuable for addressing immediate risks, they may fail to capture the broader systemic and cross-border implications of climate-induced disruptions, particularly within interconnected global supply chains. Few studies adopt a global or multi-scale approach, limiting insights into how regional disruptions cascade through international trade networks. Furthermore, the emphasis on short- to medium-term time horizons in many studies contrasts sharply with the need for long-term climate adaptation planning. Long-term risks, such as the gradual effects of sea-level rise and recurring extreme weather events, require forward-looking strategies that extend beyond immediate, reactive measures. Without this perspective, ports risk being unprepared for cumulative and compounding impacts over time.\\
\indent From a methodological perspective, risk assessments, and adaptation planning dominate the literature, reflecting the critical need to evaluate vulnerabilities and design strategic responses to climate hazards. However, methodological diversity remains limited. Many studies rely on descriptive approaches such as case studies and empirical research using surveys and interviews. While these methods offer valuable context-specific insights, the limited adoption of advanced, technology-driven approaches highlights a significant missed opportunity. Tools such as digital twinning, advanced simulation models, and AI-driven analytics appear in only a handful of recent studies, despite their potential to provide real-time, predictive insights that can dynamically inform decision-making. For instance, digital twins could simulate the cascading effects of climate-induced disruptions across interconnected systems, enabling stakeholders to test adaptive strategies before implementation. Similarly, AI-driven analytics can process complex datasets to uncover emerging vulnerabilities and optimize resource allocation.\\
\indent Additionally, studies incorporating supply chain resilience remain sparse, indicating a lack of comprehensive, end-to-end investigations that extend beyond port infrastructure to the broader logistics network. This omission is significant because supply chain resilience is critical to ensuring the continuity of global trade during and after climate disruptions. A more integrated approach that bridges the gap between infrastructure, operations, and supply chain systems could lead to holistic solutions, enhancing resilience across all dimensions of the maritime network.\\
\indent Overall, the table shows that while the existing literature has made substantial progress in identifying risks and proposing adaptation strategies, several critical gaps persist. Future studies should aim to incorporate underexplored disruptions like drought, which are increasingly relevant in the context of water scarcity and its effects on inland navigation. Moreover, expanding the use of innovative tools such as digital twins, advanced simulations, and AI can significantly improve the capacity to predict and manage climate impacts dynamically. Adopting a broader, multi-scale perspective that addresses the interconnectedness of infrastructure, operations, and supply chains is essential for crafting effective, forward-looking resilience strategies. By addressing these gaps, the field can move beyond localized, short-term solutions to develop more robust and adaptive systems capable of navigating the complex challenges posed by climate change.
\section{Existing Challenges and Pathways Ahead}
\label{Dis}
\noindent The analysis of the literature on climate-resilient port and waterborne transport systems reveals significant progress while also identifying critical gaps and challenges that must be addressed to enhance the resilience of these systems. These challenges span methodological, technological, and thematic dimensions, often limiting the ability to respond effectively to climate-induced disruptions. To overcome these gaps, pathways for future research and practice must emphasize integration, innovation, and collaboration to ensure sustainable and resilient outcomes.
\subsection{Limited Focus on Specific Climate-Induced Disruptions}
\noindent A critical challenge in the current body of research is the tendency to adopt a generalized perspective on climate-induced disruptions, often overlooking the specific impacts of individual disruption types. Of the 46 papers reviewed, a significant portion (30) fails to address particular disruptions, such as sea-level rise, flooding, drought, hurricanes, or extreme wind events. While broad analyses are valuable for identifying overarching trends, tailored strategies are essential for addressing the unique impacts and management requirements of each disruption type. For example, sea-level rise necessitates long-term infrastructural adaptations, such as constructing modular infrastructure and integrating adaptive technologies like flood-proof port designs. In contrast, flooding-related disruptions call for operational adjustments, including enhanced drainage systems and the optimization of inland waterway navigation protocols during extreme weather events. \\
\indent The underrepresentation of drought as a specific disruption in existing studies is particularly concerning. Water scarcity poses critical challenges to inland waterways and vessel operations, with potentially severe economic consequences. This was starkly illustrated during the severe droughts along the Rhine River in 2018 and 2022, which drastically reduced water levels, limiting the navigability of the river. Cargo ships were forced to operate at significantly reduced capacities or suspend operations altogether, resulting in considerable economic losses and logistical bottlenecks. These events demonstrate not only the direct operational impacts of drought but also its cascading effects on regional and international supply chains, further emphasizing the interconnected vulnerabilities of port systems.\\
\indent Adding to the complexity, many climate-induced disruptions are likely to occur concurrently or sequentially, compounding their impacts and complicating resilience strategies. For instance, during inland waterway droughts, the potential for wildfires in surrounding regions may escalate due to dry conditions, further limiting alternative transport options like roadways. Similarly, in coastal areas, the simultaneous occurrence of hurricanes and flooding can overwhelm infrastructure and disrupt both port operations and hinterland connections. The combination of multiple disruptions presents unique challenges, as traditional contingency plans often fail to account for their interconnected and mutually reinforcing effects. The lack of research on these compound events limits the development of robust, integrated resilience strategies.\\
\indent To bridge these gaps, future research should prioritize the study of specific disruptions and their associated challenges, including the potential for overlapping or sequential events, to develop targeted and actionable strategies. Advanced tools and methodologies, such as digital twins, hydrodynamic models, machine learning algorithms, and predictive analytics, offer the potential to simulate the unique effects of disruptions like drought and extreme weather events, as well as their compounded impacts. For instance, digital twins could provide real-time, scenario-based planning to assess the ripple effects of combined disruptions, such as drought-induced waterway limitations paired with roadway constraints caused by wildfires. Meanwhile, hydrodynamic models can enhance preparedness for flooding by predicting water flow changes, erosion risks, and flood probabilities, even in scenarios where coastal and inland disruptions occur simultaneously. The integration of adaptive water management strategies, supported by real-time data collection through IoT (Internet of Things) sensors, can further equip ports to manage complex and interconnected climate risks. By focusing on the particularities of individual disruptions and their interactions, researchers and policymakers can craft more effective, context-specific solutions to enhance resilience across port and hinterland systems.
\subsection{Overemphasis on Infrastructure with Limited Supply Chain Focus}
\begin{figure}[h]
\centering
    \includegraphics[ clip, trim=0cm 6cm 0cm 0.3cm, width=1\textwidth]{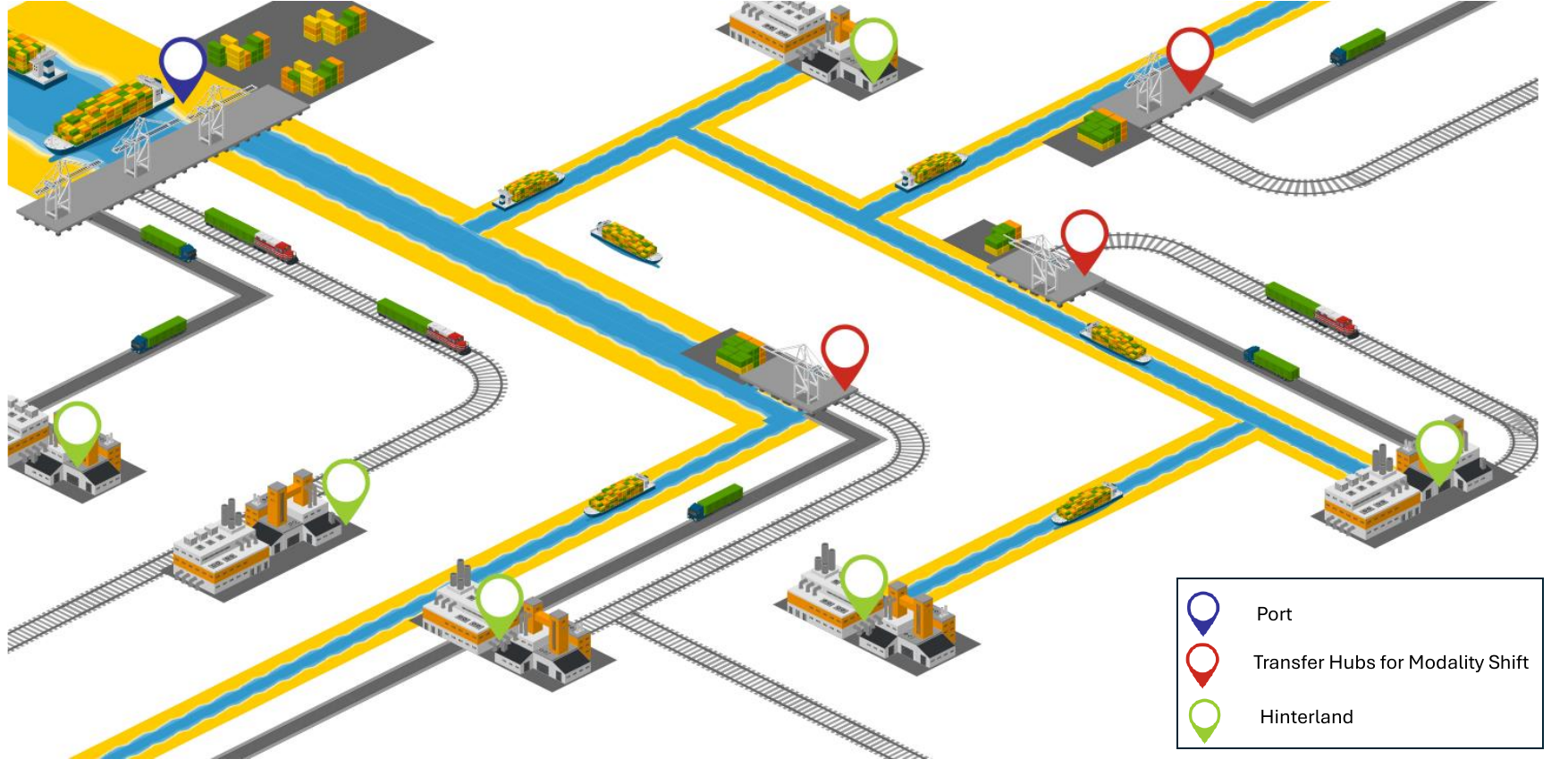}
    \caption{Multi-Modal Port-Hinterland Connections}
    \label{Fig-modality_network}
\end{figure}

\noindent The majority of existing studies predominantly focus on port infrastructure, with 41\% addressing infrastructure-related challenges compared to 39\% that concentrate on operational issues. In contrast, only 15\% of studies emphasize supply chain resilience. This uneven distribution of focus underscores a critical gap in comprehending the complex, interconnected nature of the port-hinterland system. Ports do not function in isolation; rather, they are integral nodes within a broader network where disruptions in infrastructure and operations can trigger cascading effects. These disruptions ripple through logistics chains, transportation networks, and economic systems, often amplifying their impacts beyond the immediate vicinity of the port.\\
\indent By predominantly addressing individual components, such as infrastructure or operations, research overlooks the interconnected vulnerabilities of the system and the potential for integrated resilience strategies that encompass the full maritime network. Furthermore, the literature largely neglects the potential for strategic "coopetition" among ports, a collaboration between ports that may traditionally view each other as competitors. Such collaboration could lead to shared operational efficiencies and bolster the resilience of the entire network. For instance, ports could jointly invest in regional disaster response systems or share data on cargo movements to optimize resource allocation during disruptions. This aspect of port relationships remains underexplored, further constraining the development of comprehensive strategies. Similarly, a supply chain-wide perspective can reveal opportunities for modality shifts across the network, offering innovative solutions in the face of climate-induced disruptions. For example, during a drought that renders inland waterways costly or infeasible, a system-wide approach could identify transfer hubs where cargo can shift from water to rail or road transportation. Such hubs, strategically located within the network, enable the continuity of supply chain operations while minimizing delays and additional costs. This modality shift not only enhances resilience but also ensures efficient resource utilization. Figure~\ref{Fig-modality_network} illustrates a network of modalities from port to hinterland, highlighting transfer hubs where shifts between water, rail, and road can occur to adapt to disruptions.\\
\indent In response, future research should adopt a more holistic perspective that encompasses infrastructure, operations, and the broader supply chain. A systems-based approach is essential for identifying interdependencies and crafting comprehensive strategies to mitigate cascading risks. Collaborative frameworks that include not only ports but also inland logistics providers, transportation stakeholders, and other key actors are particularly vital. Such frameworks could also facilitate multi-stakeholder decision-making platforms to align priorities, reduce redundancies, and streamline adaptation measures. Within this context, strategic coopetition among ports represents an untapped opportunity to enhance operational efficiency and strengthen collective resilience. Advanced analytical tools, such as Game Theory, Agent-Based Modeling, and Multi-Criteria Decision Analysis (MCDA), offer significant potential for evaluating trade-offs among competing objectives and devising coordinated adaptation pathways. Furthermore, the integration of blockchain technology can play a transformative role by improving transparency across the supply chain. Blockchain can enable secure, real-time data sharing on port performance metrics, vessel schedules, and disruption responses, fostering trust among stakeholders while enhancing risk management capabilities.
\subsection{Methodological Gaps and Limited Use of Advanced Technologies}
\noindent From a methodological perspective, the current literature on climate resilience in port and supply chain systems is heavily dominated by risk assessments and case studies, which primarily rely on descriptive and qualitative approaches. While these methods are invaluable for providing foundational insights and contextual understanding, they often lack the predictive power and dynamic adaptability required to address complex, evolving climate challenges effectively. The underutilization of advanced technologies and quantitative models constrains the field's ability to predict, simulate, and manage climate impacts in a proactive and data-driven manner. Furthermore, empirical and experimental research, critical for validating theoretical models and bridging the gap between conceptual frameworks and practical applications, remains sparse. The lack of empirical validation weakens the ability of current research to translate theoretical insights into actionable, field-tested solutions.\\
\indent Notably, emerging tools such as AI, digital twins, and advanced simulation technologies have only recently begun to appear in a small subset of studies, leaving their potential largely untapped. For example, while digital twins have been used in other infrastructure sectors to great effect, their application in ports and inland waterways remains limited to pilot projects, often lacking scalability. They could go beyond their current applications to create virtual ships and infrastructure that interact with the environment, simulating the performance of vessel fleets under varying climate conditions. This could also include simulating responses to changes in inland waterway conditions, such as low water levels during droughts. Similarly, creating virtual captains using reinforcement learning (RL) could provide insights into decision-making under stress, mimicking human behavior during extreme weather events or disruptions. These capabilities could significantly improve the applicability of climate impact simulations. \\
\indent Reconstructions of past shipping incidents using data fusion techniques could provide valuable insights into vulnerabilities within port-hinterland systems. For instance, by analyzing historical data on incidents caused by extreme weather or infrastructure failures, researchers can identify patterns and inform the development of preemptive measures. Additionally, detecting changes in sea or inland waterway infrastructure using satellite imagery offers a powerful tool for monitoring vulnerabilities. These methods can provide near-real-time updates on infrastructure conditions, supporting proactive maintenance and risk mitigation. Additionally, techniques such as detecting anomalous behavior on maritime routes, through embedding and manifold methods, could provide early warnings of disruptions or unexpected vessel activity. This would be particularly relevant for ensuring continuity in port-hinterland logistics under shifting environmental or operational conditions.\\
\indent To address these shortcomings, it is essential for future research to embrace and integrate technological advancements that enable the creation of more dynamic and predictive tools for managing climate risks. Digital twins, for instance, hold significant promise by offering real-time virtual replicas of ports, inland waterways, and their interconnected systems. These virtual models can simulate the impacts of various climate-induced disruptions, such as extreme weather events or sea-level rise, allowing stakeholders to test adaptive strategies under controlled conditions before implementation. Similarly, AI-driven analytics have the capacity to process vast and complex datasets, uncover hidden patterns, forecast risks, and provide actionable insights for decision-making processes. Combined with simulation tools and hydrodynamic models, these technologies can map the cascading effects of disruptions across infrastructure, operations, and supply chains, enabling a more comprehensive understanding of systemic vulnerabilities.\\
\indent Equally important is the integration of empirical methods to ground these technological innovations in practical, real-world conditions. Stakeholder engagement through interviews, surveys, and workshops can provide invaluable feedback on the applicability and feasibility of proposed tools and strategies. Participatory modeling approaches, where stakeholders actively contribute to the design of models, can further enhance the relevance and acceptability of these technologies. Field observations and longitudinal studies can validate the predictive capabilities of digital twins, AI models, and simulations, ensuring that theoretical advancements translate into effective and sustainable solutions.\\ 
\indent By incorporating advanced technologies such as data fusion, satellite imagery, reinforcement learning, and anomaly detection into empirical research frameworks, the field can move toward a more holistic, evidence-based approach to climate resilience. These tools and methods can significantly enhance the capacity of port-hinterland systems to anticipate, respond to, and recover from climate-induced disruptions.
\subsection{Uneven Geographic Distribution and Short-Term Focus}
\noindent The analysis reveals significant disparities in the geographic distribution of research on climate resilience in ports. Leading port nations, such as the United States and the Netherlands, exhibit high levels of research activity. In contrast, countries with equally critical ports, such as China, Singapore, and South Korea, contribute comparatively little to the literature. This uneven distribution hinders a comprehensive global understanding of climate resilience, particularly in regions highly vulnerable to climate change and central to global trade networks. At the same time, much of the existing research focuses on short- to medium-term time horizons, neglecting the long-term nature of climate disruptions like sea-level rise, which evolve gradually but with compounding impacts over decades. This narrow temporal focus leaves ports and inland waterways susceptible to cumulative operational and economic risks in the future. \\
\indent Addressing these challenges requires fostering global collaboration in climate resilience research and extending planning frameworks to include long-term perspectives. Promoting resilience-focused studies in underrepresented regions is essential to bridge the global research gap. International initiatives such as the Clarion and SAFARI projects in the EU provide exemplary models by integrating advanced technologies, including UAVs, sensor networks, and digital twins, to enhance port climate resilience. Encouraging partnerships between academic institutions, governments, and industry stakeholders can drive research in underrepresented regions, while knowledge-sharing platforms can disseminate findings and best practices on a global scale.\\
\indent To ensure sustainable, long-term resilience, research and planning must integrate tools such as scenario-based modeling and adaptation pathways. These approaches allow stakeholders to project future climate impacts, assess uncertainties, and implement phased, flexible responses. For example, adaptation pathways, as demonstrated by \citet{brooke2024port}, enable ports to adopt incremental measures while retaining the agility to respond to evolving risks. Additionally, strategic investments in green infrastructure, such as mangroves and vegetated dunes, can provide durable protection against hazards like flooding and erosion, combining ecological benefits with long-term climate resilience. By addressing both geographic and temporal gaps, the field can better equip ports and associated systems to withstand the multifaceted challenges posed by climate change.
\section{Conclusion}
\label{Conc}
\noindent \noindent This study provides a comprehensive review of the current state of research on climate resilience in port and waterborne transport systems, offering critical insights into existing gaps and opportunities for advancement. The findings reveal several pressing challenges: an overemphasis on port infrastructure (41\%) compared to supply chain resilience (15\%), a limited focus on specific climate-induced disruptions such as drought and compounded events like simultaneous drought and wildfires, and a tendency toward short- to medium-term planning that neglects long-term adaptation needs. These gaps are further accompanied by geographic disparities in research output, with significant underrepresentation from regions hosting critical ports, and by methodological constraints, where risk assessments and case studies dominate while advanced technologies remain underutilized.\\
\indent Emerging tools like digital twins, artificial intelligence, and satellite monitoring hold significant potential to address these limitations, yet their application in the field is still in its infancy. For instance, digital twins can simulate complex scenarios, including cascading disruptions, while AI and data fusion can uncover vulnerabilities and inform real-time decision-making. Furthermore, the lack of research on the interplay of simultaneous disruptions highlights an urgent need to explore these compounded risks in future studies. To bridge these gaps, this study emphasizes the importance of adopting systems-based approaches that integrate infrastructure, operations, and supply chains. Collaborative frameworks involving diverse stakeholders, supported by participatory modeling and empirical validation, are essential to developing adaptive strategies that account for cascading risks and systemic vulnerabilities.\\
\indent By addressing these challenges, this paper contributes to a forward-looking research agenda aimed at bolstering climate resilience in waterborne transport systems. It underscores the transformative role of technological innovation and holistic frameworks in ensuring sustainable and adaptive infrastructure capable of withstanding the multifaceted impacts of climate change.
 \\

\section*{Acknowledgment} 
\noindent Funded by the European Union. Views and opinions expressed are, however, those of the author(s) only and do not necessarily reflect those of the European Union or the Horizon Europe research and innovation programme. Neither the European Union nor the granting authority can be held responsible for them.\\
\bibliographystyle{elsarticle-harv} 
\bibliography{00-Paper}

\begin{thebibliography}{74}
\expandafter\ifx\csname natexlab\endcsname\relax\def\natexlab#1{#1}\fi
\providecommand{\url}[1]{\texttt{#1}}
\providecommand{\href}[2]{#2}
\providecommand{\path}[1]{#1}
\providecommand{\DOIprefix}{doi:}
\providecommand{\ArXivprefix}{arXiv:}
\providecommand{\URLprefix}{URL: }
\providecommand{\Pubmedprefix}{pmid:}
\providecommand{\doi}[1]{\href{http://dx.doi.org/#1}{\path{#1}}}
\providecommand{\Pubmed}[1]{\href{pmid:#1}{\path{#1}}}
\providecommand{\bibinfo}[2]{#2}
\ifx\xfnm\relax \def\xfnm[#1]{\unskip,\space#1}\fi
\bibitem[{Abdelhafez et~al.(2021)Abdelhafez, Ellingwood and Mahmoud}]{abdelhafez2021vulnerability}
\bibinfo{author}{Abdelhafez, M.A.}, \bibinfo{author}{Ellingwood, B.}, \bibinfo{author}{Mahmoud, H.}, \bibinfo{year}{2021}.
\newblock \bibinfo{title}{Vulnerability of seaports to hurricanes and sea level rise in a changing climate: A case study for mobile, al}.
\newblock \bibinfo{journal}{Coastal Engineering} \bibinfo{volume}{167}, \bibinfo{pages}{103884}.
\bibitem[{Ahmadi et~al.(2025)Ahmadi, Lather, Wittich and Madson}]{ahmadi2025current}
\bibinfo{author}{Ahmadi, S.Y.}, \bibinfo{author}{Lather, J.I.}, \bibinfo{author}{Wittich, C.E.}, \bibinfo{author}{Madson, K.}, \bibinfo{year}{2025}.
\newblock \bibinfo{title}{Current and future trends for resilient inland waterway transportation systems during flood disruptions}.
\newblock \bibinfo{journal}{Journal of Construction Engineering and Management} \bibinfo{volume}{151}, \bibinfo{pages}{04025046}.
\bibitem[{Allen et~al.(2022)Allen, McLeod and Hutt}]{allen2022sea}
\bibinfo{author}{Allen, T.R.}, \bibinfo{author}{McLeod, G.}, \bibinfo{author}{Hutt, S.}, \bibinfo{year}{2022}.
\newblock \bibinfo{title}{Sea level rise exposure assessment of us east coast cargo container terminals}.
\newblock \bibinfo{journal}{Maritime Policy \& Management} \bibinfo{volume}{49}, \bibinfo{pages}{577--599}.
\bibitem[{Asadabadi and M-Hooks(2020)}]{asadabadi2020maritime}
\bibinfo{author}{Asadabadi, A.}, \bibinfo{author}{M-Hooks, E.}, \bibinfo{year}{2020}.
\newblock \bibinfo{title}{Maritime port network resiliency and reliability through co-opetition}.
\newblock \bibinfo{journal}{Transportation Research Part E: Logistics and Transportation Review} \bibinfo{volume}{137}, \bibinfo{pages}{101916}.
\bibitem[{Bakker and van Koningsveld(2024)}]{Bakker2024PAINC}
\bibinfo{author}{Bakker, F.}, \bibinfo{author}{van Koningsveld, M.}, \bibinfo{year}{2024}.
\newblock \bibinfo{title}{Tool to evaluate countermeasures at locks to limit freshwater losses and saltwater intrusion while minimizing waiting times for shipping}, in: \bibinfo{booktitle}{Proceedings of the 35th PIANC World Congress, 29 April - 3 May 2024, Cape Town, South Africa}.
\bibitem[{Bal and Vleugel(2023)}]{bal2023towards}
\bibinfo{author}{Bal, F.}, \bibinfo{author}{Vleugel, J.}, \bibinfo{year}{2023}.
\newblock \bibinfo{title}{Towards climate resilient freight transport in europe}.
\newblock \bibinfo{journal}{International Journal of Transport Development and Integration} \bibinfo{volume}{7}, \bibinfo{pages}{147--152}.
\bibitem[{Balakrishnan et~al.(2022)Balakrishnan, Lim and Zhang}]{balakrishnan2022methodology}
\bibinfo{author}{Balakrishnan, S.}, \bibinfo{author}{Lim, T.}, \bibinfo{author}{Zhang, Z.}, \bibinfo{year}{2022}.
\newblock \bibinfo{title}{A methodology for evaluating the economic risks of hurricane-related disruptions to port operations}.
\newblock \bibinfo{journal}{Transportation Research Part A: Policy and Practice} \bibinfo{volume}{162}, \bibinfo{pages}{58--79}.
\bibitem[{Becker and Caldwell(2015)}]{becker2015stakeholder}
\bibinfo{author}{Becker, A.}, \bibinfo{author}{Caldwell, M.R.}, \bibinfo{year}{2015}.
\newblock \bibinfo{title}{Stakeholder perceptions of seaport resilience strategies: A case study of gulfport (mississippi) and providence (rhode island)}.
\newblock \bibinfo{journal}{Coastal Management} \bibinfo{volume}{43}, \bibinfo{pages}{1--34}.
\bibitem[{Becker et~al.(2016)Becker, Chase, Fischer, Schwegler and Mosher}]{becker2016method}
\bibinfo{author}{Becker, A.}, \bibinfo{author}{Chase, N.T.}, \bibinfo{author}{Fischer, M.}, \bibinfo{author}{Schwegler, B.}, \bibinfo{author}{Mosher, K.}, \bibinfo{year}{2016}.
\newblock \bibinfo{title}{A method to estimate climate-critical construction materials applied to seaport protection}.
\newblock \bibinfo{journal}{Global environmental change} \bibinfo{volume}{40}, \bibinfo{pages}{125--136}.
\bibitem[{Brooke et~al.(2024)Brooke, Giovando, Dahl, Joyner and Herbert}]{brooke2024port}
\bibinfo{author}{Brooke, J.}, \bibinfo{author}{Giovando, J.}, \bibinfo{author}{Dahl, T.A.}, \bibinfo{author}{Joyner, B.}, \bibinfo{author}{Herbert, L.}, \bibinfo{year}{2024}.
\newblock \bibinfo{title}{Port and inland waterway design and operation in the face of climate change uncertainty}, in: \bibinfo{booktitle}{Proceedings of the Institution of Civil Engineers-Civil Engineering}, \bibinfo{organization}{Emerald Publishing Limited}. pp. \bibinfo{pages}{37--49}.
\bibitem[{Chowdhooree et~al.(2024)Chowdhooree, Aziz, Rashid and Hossain}]{chowdhooree2024climate}
\bibinfo{author}{Chowdhooree, I.}, \bibinfo{author}{Aziz, T.}, \bibinfo{author}{Rashid, M.J.}, \bibinfo{author}{Hossain, M.}, \bibinfo{year}{2024}.
\newblock \bibinfo{title}{Climate change adaptation through nature-based solution: examining the case of thakurani khal of mongla port municipality, bagerhat bangladesh}.
\newblock \bibinfo{journal}{International Journal of Disaster Resilience in the Built Environment} \bibinfo{volume}{15}, \bibinfo{pages}{474--493}.
\bibitem[{Christodoulou et~al.(2019)Christodoulou, Christidis and Demirel}]{christodoulou2019sea}
\bibinfo{author}{Christodoulou, A.}, \bibinfo{author}{Christidis, P.}, \bibinfo{author}{Demirel, H.}, \bibinfo{year}{2019}.
\newblock \bibinfo{title}{Sea-level rise in ports: a wider focus on impacts}.
\newblock \bibinfo{journal}{Maritime Economics \& Logistics} \bibinfo{volume}{21}, \bibinfo{pages}{482--496}.
\bibitem[{CLARION(2024-2028)}]{Clarion}
\bibinfo{author}{CLARION}, \bibinfo{year}{2024-2028}.
\newblock \bibinfo{title}{Climate resilient port infrastructure} .
\bibitem[{Cristal(2023-2025)}]{Cristal}
\bibinfo{author}{Cristal}, \bibinfo{year}{2023-2025}.
\newblock \bibinfo{title}{Climate resilient and environmentally sustainable transport infrastructure} .
\bibitem[{Deltares(2023)}]{Deltares2023}
\bibinfo{author}{Deltares}, \bibinfo{year}{2023}.
\newblock \bibinfo{title}{Digital twins promising tools for water and subsurface management}.
\newblock \bibinfo{journal}{The Deltares perspective} .
\bibitem[{Dircke and Molenaar(2015)}]{dircke2015climate}
\bibinfo{author}{Dircke, P.}, \bibinfo{author}{Molenaar, A.}, \bibinfo{year}{2015}.
\newblock \bibinfo{title}{Climate change adaptation; innovative tools and strategies in delta city rotterdam}.
\newblock \bibinfo{journal}{Water Practice and Technology} \bibinfo{volume}{10}, \bibinfo{pages}{674--680}.
\bibitem[{van Dorsser et~al.(2020)van Dorsser, Vinke, Hekkenberg and van Koningsveld}]{van2020effect}
\bibinfo{author}{van Dorsser, C.}, \bibinfo{author}{Vinke, F.}, \bibinfo{author}{Hekkenberg, R.}, \bibinfo{author}{van Koningsveld, M.}, \bibinfo{year}{2020}.
\newblock \bibinfo{title}{The effect of low water on loading capacity of inland ships}.
\newblock \bibinfo{journal}{European Journal of Transport and Infrastructure Research} \bibinfo{volume}{20}, \bibinfo{pages}{47--70}.
\bibitem[{Echendu(2021)}]{echendu2021relationship}
\bibinfo{author}{Echendu, A.J.}, \bibinfo{year}{2021}.
\newblock \bibinfo{title}{Relationship between urban planning and flooding in port harcourt city, nigeria; insights from planning professionals}.
\newblock \bibinfo{journal}{Journal of Flood Risk Management} \bibinfo{volume}{14}, \bibinfo{pages}{e12693}.
\bibitem[{Folkman et~al.(2021)Folkman, Gharehgozli, Mileski and Galvao}]{folkman2021port}
\bibinfo{author}{Folkman, D.}, \bibinfo{author}{Gharehgozli, A.}, \bibinfo{author}{Mileski, J.}, \bibinfo{author}{Galvao, C.B.}, \bibinfo{year}{2021}.
\newblock \bibinfo{title}{Port resiliency and the effects of hurricanes on port operations}.
\newblock \bibinfo{journal}{International Journal of Advanced Operations Management} \bibinfo{volume}{13}, \bibinfo{pages}{409--430}.
\bibitem[{Fouad et~al.(2023)Fouad, Heggy and Weilacher}]{fouad2023waterways}
\bibinfo{author}{Fouad, S.S.}, \bibinfo{author}{Heggy, E.}, \bibinfo{author}{Weilacher, U.}, \bibinfo{year}{2023}.
\newblock \bibinfo{title}{Waterways transformation in the vulnerable port city of alexandria}.
\newblock \bibinfo{journal}{Cities} \bibinfo{volume}{141}, \bibinfo{pages}{104426}.
\bibitem[{Grant et~al.(2024)Grant, Rhode-Barbarigos, Roy, Britton, Li, Rowe, Becker, Hope and Bello}]{grant2024no}
\bibinfo{author}{Grant, R.}, \bibinfo{author}{Rhode-Barbarigos, L.}, \bibinfo{author}{Roy, S.S.}, \bibinfo{author}{Britton, L.}, \bibinfo{author}{Li, C.}, \bibinfo{author}{Rowe, A.}, \bibinfo{author}{Becker, A.}, \bibinfo{author}{Hope, B.}, \bibinfo{author}{Bello, M.}, \bibinfo{year}{2024}.
\newblock \bibinfo{title}{No port stands alone: Portmiami and the resilience of its caribbean and mesoamerican maritime network}.
\newblock \bibinfo{journal}{Maritime Economics \& Logistics} , \bibinfo{pages}{1--24}.
\bibitem[{Gu and Liu(2023)}]{gu2023systematic}
\bibinfo{author}{Gu, B.}, \bibinfo{author}{Liu, J.}, \bibinfo{year}{2023}.
\newblock \bibinfo{title}{A systematic review of resilience in the maritime transport}.
\newblock \bibinfo{journal}{International Journal of Logistics Research and Applications} , \bibinfo{pages}{1--22}.
\bibitem[{Henry and Ramirez-Marquez(2012)}]{henry2012generic}
\bibinfo{author}{Henry, D.}, \bibinfo{author}{Ramirez-Marquez, J.E.}, \bibinfo{year}{2012}.
\newblock \bibinfo{title}{Generic metrics and quantitative approaches for system resilience as a function of time}.
\newblock \bibinfo{journal}{Reliability Engineering \& System Safety} \bibinfo{volume}{99}, \bibinfo{pages}{114--122}.
\bibitem[{Hosseini and Barker(2016)}]{hosseini2016modeling}
\bibinfo{author}{Hosseini, S.}, \bibinfo{author}{Barker, K.}, \bibinfo{year}{2016}.
\newblock \bibinfo{title}{Modeling infrastructure resilience using bayesian networks: A case study of inland waterway ports}.
\newblock \bibinfo{journal}{Computers \& Industrial Engineering} \bibinfo{volume}{93}, \bibinfo{pages}{252--266}.
\bibitem[{Islam et~al.(2023)Islam, Goerlandt, Sakalayen, Shi and Venkatesh}]{islam2023developing}
\bibinfo{author}{Islam, S.}, \bibinfo{author}{Goerlandt, F.}, \bibinfo{author}{Sakalayen, Q.M.H.}, \bibinfo{author}{Shi, Y.}, \bibinfo{author}{Venkatesh, V.}, \bibinfo{year}{2023}.
\newblock \bibinfo{title}{Developing a" disaster scenario" to prepare for the possibility of disruptions to maritime transportation serving coastal communities of vancouver island}.
\newblock \bibinfo{journal}{Marine Policy} \bibinfo{volume}{150}, \bibinfo{pages}{105531}.
\bibitem[{Izaguirre et~al.(2021)Izaguirre, Losada, Camus, Vigh and Stenek}]{izaguirre2021climate}
\bibinfo{author}{Izaguirre, C.}, \bibinfo{author}{Losada, I.J.}, \bibinfo{author}{Camus, P.}, \bibinfo{author}{Vigh, J.L.}, \bibinfo{author}{Stenek, V.}, \bibinfo{year}{2021}.
\newblock \bibinfo{title}{Climate change risk to global port operations}.
\newblock \bibinfo{journal}{Nature Climate Change} \bibinfo{volume}{11}, \bibinfo{pages}{14--20}.
\bibitem[{Jarriel(2021)}]{jarriel2021climate}
\bibinfo{author}{Jarriel, K.}, \bibinfo{year}{2021}.
\newblock \bibinfo{title}{Climate disaster and the resilience of local maritime networks: Two examples from the aegean bronze age}.
\newblock \bibinfo{journal}{Quaternary International} \bibinfo{volume}{597}, \bibinfo{pages}{118--130}.
\bibitem[{Johnson et~al.(2025)Johnson, Wehbe and Baroud}]{johnson2025flood}
\bibinfo{author}{Johnson, P.M.}, \bibinfo{author}{Wehbe, C.}, \bibinfo{author}{Baroud, H.}, \bibinfo{year}{2025}.
\newblock \bibinfo{title}{How flood-resilient port infrastructure can reduce economic impacts of climate change: a case study of the us inland waterways}.
\newblock \bibinfo{journal}{Civil Engineering and Environmental Systems} , \bibinfo{pages}{1--19}.
\bibitem[{Kim et~al.(2021)Kim, Choi and Kim}]{kim2021framework}
\bibinfo{author}{Kim, S.}, \bibinfo{author}{Choi, S.}, \bibinfo{author}{Kim, C.}, \bibinfo{year}{2021}.
\newblock \bibinfo{title}{The framework for measuring port resilience in korean port case}.
\newblock \bibinfo{journal}{Sustainability} \bibinfo{volume}{13}, \bibinfo{pages}{11883}.
\bibitem[{van Koningsveld et~al.(2023)van Koningsveld, Verheij, Taneja and de~Vriend}]{van2023ports}
\bibinfo{author}{van Koningsveld, M.}, \bibinfo{author}{Verheij, H.}, \bibinfo{author}{Taneja, P.}, \bibinfo{author}{de~Vriend, H.}, \bibinfo{year}{2023}.
\newblock \bibinfo{title}{Ports and waterways: navigating the changing world}.
\newblock \bibinfo{publisher}{TU Delft OPEN Publishing}.
\bibitem[{Kontogianni et~al.(2019)Kontogianni, Damigos, Kyrtzoglou, Tourkolias and Skourtos}]{kontogianni2019development}
\bibinfo{author}{Kontogianni, A.}, \bibinfo{author}{Damigos, D.}, \bibinfo{author}{Kyrtzoglou, T.}, \bibinfo{author}{Tourkolias, C.}, \bibinfo{author}{Skourtos, M.}, \bibinfo{year}{2019}.
\newblock \bibinfo{title}{Development of a composite climate change vulnerability index for small craft harbours}.
\newblock \bibinfo{journal}{Environmental Hazards} \bibinfo{volume}{18}, \bibinfo{pages}{173--190}.
\bibitem[{Lam and Bai(2016)}]{lam2016quality}
\bibinfo{author}{Lam, J.S.L.}, \bibinfo{author}{Bai, X.}, \bibinfo{year}{2016}.
\newblock \bibinfo{title}{A quality function deployment approach to improve maritime supply chain resilience}.
\newblock \bibinfo{journal}{Transportation Research Part E: Logistics and Transportation Review} \bibinfo{volume}{92}, \bibinfo{pages}{16--27}.
\bibitem[{Lavigne and Dupray(2023)}]{lavigne202317}
\bibinfo{author}{Lavigne, C.}, \bibinfo{author}{Dupray, S.}, \bibinfo{year}{2023}.
\newblock \bibinfo{title}{17 resilience, adaptation, and adaptability}.
\newblock \bibinfo{journal}{Maritime Ports, Supply Chains and Logistics Corridors} , \bibinfo{pages}{229}.
\bibitem[{Le{\'o}n-Mateos et~al.(2021)Le{\'o}n-Mateos, Sartal, L{\'o}pez-Manuel and Quintas}]{leon2021adapting}
\bibinfo{author}{Le{\'o}n-Mateos, F.}, \bibinfo{author}{Sartal, A.}, \bibinfo{author}{L{\'o}pez-Manuel, L.}, \bibinfo{author}{Quintas, M.A.}, \bibinfo{year}{2021}.
\newblock \bibinfo{title}{Adapting our sea ports to the challenges of climate change: Development and validation of a port resilience index}.
\newblock \bibinfo{journal}{Marine Policy} \bibinfo{volume}{130}, \bibinfo{pages}{104573}.
\bibitem[{Li et~al.(2022)Li, Asadabadi and Miller-Hooks}]{li2022enhancing}
\bibinfo{author}{Li, W.}, \bibinfo{author}{Asadabadi, A.}, \bibinfo{author}{Miller-Hooks, E.}, \bibinfo{year}{2022}.
\newblock \bibinfo{title}{Enhancing resilience through port coalitions in maritime freight networks}.
\newblock \bibinfo{journal}{Transportation Research Part A: Policy and Practice} \bibinfo{volume}{157}, \bibinfo{pages}{1--23}.
\bibitem[{Liu et~al.(2018)Liu, Tian, Huang and Yang}]{liu2018analysis}
\bibinfo{author}{Liu, H.}, \bibinfo{author}{Tian, Z.}, \bibinfo{author}{Huang, A.}, \bibinfo{author}{Yang, Z.}, \bibinfo{year}{2018}.
\newblock \bibinfo{title}{Analysis of vulnerabilities in maritime supply chains}.
\newblock \bibinfo{journal}{Reliability Engineering \& System Safety} \bibinfo{volume}{169}, \bibinfo{pages}{475--484}.
\bibitem[{Liu et~al.(2023)Liu, Yang, Ng and Jiang}]{liu2023analysis}
\bibinfo{author}{Liu, Q.}, \bibinfo{author}{Yang, Y.}, \bibinfo{author}{Ng, A.K.}, \bibinfo{author}{Jiang, C.}, \bibinfo{year}{2023}.
\newblock \bibinfo{title}{An analysis on the resilience of the european port network}.
\newblock \bibinfo{journal}{Transportation Research Part A: Policy and Practice} \bibinfo{volume}{175}, \bibinfo{pages}{103778}.
\bibitem[{Mclean and Becker(2021)}]{mclean2021advancing}
\bibinfo{author}{Mclean, E.L.}, \bibinfo{author}{Becker, A.}, \bibinfo{year}{2021}.
\newblock \bibinfo{title}{Advancing seaport resilience to natural hazards due to climate change: strategies to overcome decision making barriers}.
\newblock \bibinfo{journal}{Frontiers in Sustainability} \bibinfo{volume}{2}, \bibinfo{pages}{673630}.
\bibitem[{Moher et~al.(2015)Moher, Shamseer, Clarke, Ghersi, Liberati, Petticrew, Shekelle, Stewart and Group}]{moher2015preferred}
\bibinfo{author}{Moher, D.}, \bibinfo{author}{Shamseer, L.}, \bibinfo{author}{Clarke, M.}, \bibinfo{author}{Ghersi, D.}, \bibinfo{author}{Liberati, A.}, \bibinfo{author}{Petticrew, M.}, \bibinfo{author}{Shekelle, P.}, \bibinfo{author}{Stewart, L.A.}, \bibinfo{author}{Group, P.P.}, \bibinfo{year}{2015}.
\newblock \bibinfo{title}{Preferred reporting items for systematic review and meta-analysis protocols (prisma-p) 2015 statement}.
\newblock \bibinfo{journal}{Systematic reviews} \bibinfo{volume}{4}, \bibinfo{pages}{1--9}.
\bibitem[{Moreno-Fern{\'a}ndez et~al.(2021)Moreno-Fern{\'a}ndez, Zavala, Madrigal-Gonz{\'a}lez and Seijo}]{moreno2021resilience}
\bibinfo{author}{Moreno-Fern{\'a}ndez, D.}, \bibinfo{author}{Zavala, M.A.}, \bibinfo{author}{Madrigal-Gonz{\'a}lez, J.}, \bibinfo{author}{Seijo, F.}, \bibinfo{year}{2021}.
\newblock \bibinfo{title}{Resilience as a moving target: An evaluation of last century management strategies in a dry-edge maritime pine ecosystem}.
\newblock \bibinfo{journal}{Forests} \bibinfo{volume}{12}, \bibinfo{pages}{1151}.
\bibitem[{Morris(2020)}]{morris2020stakeholder}
\bibinfo{author}{Morris, L.L.}, \bibinfo{year}{2020}.
\newblock \bibinfo{title}{Stakeholder collaboration as a pathway to climate adaptation at coastal ports}.
\newblock \bibinfo{journal}{Maritime Policy \& Management} \bibinfo{volume}{47}, \bibinfo{pages}{953--967}.
\bibitem[{{National Infrastructure Advisory Council (NIAC)}(2010)}]{resilience2010national}
\bibinfo{author}{{National Infrastructure Advisory Council (NIAC)}}, \bibinfo{year}{2010}.
\newblock \bibinfo{title}{Resilience, infrastructure} .
\bibitem[{Neumann et~al.(2015)Neumann, Vafeidis, Zimmermann and Nicholls}]{neumann2015future}
\bibinfo{author}{Neumann, B.}, \bibinfo{author}{Vafeidis, A.T.}, \bibinfo{author}{Zimmermann, J.}, \bibinfo{author}{Nicholls, R.J.}, \bibinfo{year}{2015}.
\newblock \bibinfo{title}{Future coastal population growth and exposure to sea-level rise and coastal flooding-a global assessment}.
\newblock \bibinfo{journal}{PloS one} \bibinfo{volume}{10}, \bibinfo{pages}{e0118571}.
\bibitem[{Nguyen et~al.(2023)Nguyen, My~Tran, Duc and Thai}]{nguyen2023managing}
\bibinfo{author}{Nguyen, T.T.}, \bibinfo{author}{My~Tran, D.T.}, \bibinfo{author}{Duc, T.T.H.}, \bibinfo{author}{Thai, V.V.}, \bibinfo{year}{2023}.
\newblock \bibinfo{title}{Managing disruptions in the maritime industry--a systematic literature review}.
\newblock \bibinfo{journal}{Maritime Business Review} \bibinfo{volume}{8}, \bibinfo{pages}{170--190}.
\bibitem[{Polydoropoulou et~al.(2025)Polydoropoulou, Bouhouras, Papaioannou and Karakikes}]{polydoropoulou2025living}
\bibinfo{author}{Polydoropoulou, A.}, \bibinfo{author}{Bouhouras, E.}, \bibinfo{author}{Papaioannou, G.}, \bibinfo{author}{Karakikes, I.}, \bibinfo{year}{2025}.
\newblock \bibinfo{title}{Living labs for the resilience of ports against climate change disruptions}.
\newblock \bibinfo{journal}{Ocean \& Coastal Management} \bibinfo{volume}{261}, \bibinfo{pages}{107528}.
\bibitem[{Ponomarov and Holcomb(2009)}]{ponomarov2009understanding}
\bibinfo{author}{Ponomarov, S.Y.}, \bibinfo{author}{Holcomb, M.C.}, \bibinfo{year}{2009}.
\newblock \bibinfo{title}{Understanding the concept of supply chain resilience}.
\newblock \bibinfo{journal}{The international journal of logistics management} \bibinfo{volume}{20}, \bibinfo{pages}{124--143}.
\bibitem[{Poo and Yang(2024)}]{poo2024optimising}
\bibinfo{author}{Poo, M.C.P.}, \bibinfo{author}{Yang, Z.}, \bibinfo{year}{2024}.
\newblock \bibinfo{title}{Optimising the resilience of shipping networks to climate vulnerability}.
\newblock \bibinfo{journal}{Maritime Policy \& Management} \bibinfo{volume}{51}, \bibinfo{pages}{15--34}.
\bibitem[{Poo et~al.(2021)Poo, Yang, Dimitriu, Qu, Jin and Feng}]{poo2021climate}
\bibinfo{author}{Poo, M.C.P.}, \bibinfo{author}{Yang, Z.}, \bibinfo{author}{Dimitriu, D.}, \bibinfo{author}{Qu, Z.}, \bibinfo{author}{Jin, Z.}, \bibinfo{author}{Feng, X.}, \bibinfo{year}{2021}.
\newblock \bibinfo{title}{Climate change risk indicators (ccri) for seaports in the united kingdom}.
\newblock \bibinfo{journal}{Ocean \& Coastal Management} \bibinfo{volume}{205}, \bibinfo{pages}{105580}.
\bibitem[{ReNEW(2023-2025)}]{ReNEW}
\bibinfo{author}{ReNEW}, \bibinfo{year}{2023-2025}.
\newblock \bibinfo{title}{Resilience-centric smart, green, networked eu inland waterways} .
\bibitem[{Repetto et~al.(2017)Repetto, Burlando, Solari, De~Gaetano and Pizzo}]{repetto2017integrated}
\bibinfo{author}{Repetto, M.P.}, \bibinfo{author}{Burlando, M.}, \bibinfo{author}{Solari, G.}, \bibinfo{author}{De~Gaetano, P.}, \bibinfo{author}{Pizzo, M.}, \bibinfo{year}{2017}.
\newblock \bibinfo{title}{Integrated tools for improving the resilience of seaports under extreme wind events}.
\newblock \bibinfo{journal}{Sustainable cities and society} \bibinfo{volume}{32}, \bibinfo{pages}{277--294}.
\bibitem[{Ribeiro et~al.(2021)Ribeiro, Lopes, Sousa, Gomez-Gesteira and Dias}]{ribeiro2021flooding}
\bibinfo{author}{Ribeiro, A.S.}, \bibinfo{author}{Lopes, C.L.}, \bibinfo{author}{Sousa, M.C.}, \bibinfo{author}{Gomez-Gesteira, M.}, \bibinfo{author}{Dias, J.M.}, \bibinfo{year}{2021}.
\newblock \bibinfo{title}{Flooding conditions at aveiro port (portugal) within the framework of projected climate change}.
\newblock \bibinfo{journal}{Journal of Marine Science and Engineering} \bibinfo{volume}{9}, \bibinfo{pages}{595}.
\bibitem[{Romya et~al.(2025)Romya, Elkut, Zed, Moghazy, El-Tahhan, Soliman, Ibrahim, Iskander and Sharaan}]{romya2025adaptation}
\bibinfo{author}{Romya, A.A.}, \bibinfo{author}{Elkut, A.E.}, \bibinfo{author}{Zed, A.A.A.}, \bibinfo{author}{Moghazy, H.M.}, \bibinfo{author}{El-Tahhan, M.K.}, \bibinfo{author}{Soliman, A.}, \bibinfo{author}{Ibrahim, M.G.}, \bibinfo{author}{Iskander, M.}, \bibinfo{author}{Sharaan, M.}, \bibinfo{year}{2025}.
\newblock \bibinfo{title}{Adaptation and development plans of the egyptian ports under the impacts of climate change}.
\newblock \bibinfo{journal}{Ocean \& Coastal Management} \bibinfo{volume}{262}, \bibinfo{pages}{107577}.
\bibitem[{Rubinstein et~al.(2023)Rubinstein, Machado, Sharma, Monkiewicz and Morris}]{rubinstein2023nature}
\bibinfo{author}{Rubinstein, T.S.}, \bibinfo{author}{Machado, D.A.}, \bibinfo{author}{Sharma, R.}, \bibinfo{author}{Monkiewicz, S.J.}, \bibinfo{author}{Morris, R.L.}, \bibinfo{year}{2023}.
\newblock \bibinfo{title}{Nature based method coastal hazard adaptation options-site selection in western port, victoria}, in: \bibinfo{booktitle}{Australasian Coasts \& Ports 2023 Conference}, \bibinfo{organization}{Engineers Australia Sunshine Coast, QLD}. pp. \bibinfo{pages}{718--727}.
\bibitem[{Ryan-Henry and Becker(2020)}]{ryan2020port}
\bibinfo{author}{Ryan-Henry, J.}, \bibinfo{author}{Becker, A.}, \bibinfo{year}{2020}.
\newblock \bibinfo{title}{Port stakeholder perceptions of sandy impacts: A case study of red hook, new york}.
\newblock \bibinfo{journal}{Maritime Policy \& Management} \bibinfo{volume}{47}, \bibinfo{pages}{885--902}.
\bibitem[{{SAFARI Ports}(2024-2027)}]{Safari}
\bibinfo{author}{{SAFARI Ports}}, \bibinfo{year}{2024-2027}.
\newblock \bibinfo{title}{Safe and climate resilient ports} .
\bibitem[{Salem et~al.(2020)Salem, Siam, El-Dakhakhni and Tait}]{salem2020probabilistic}
\bibinfo{author}{Salem, S.}, \bibinfo{author}{Siam, A.}, \bibinfo{author}{El-Dakhakhni, W.}, \bibinfo{author}{Tait, M.}, \bibinfo{year}{2020}.
\newblock \bibinfo{title}{Probabilistic resilience-guided infrastructure risk management}.
\newblock \bibinfo{journal}{Journal of Management in Engineering} \bibinfo{volume}{36}, \bibinfo{pages}{04020073}.
\bibitem[{Saswat et~al.(2024)Saswat, Seetharaman, Maddulety and Bakhshi}]{saswat2024adoption}
\bibinfo{author}{Saswat, S.S.}, \bibinfo{author}{Seetharaman, A.}, \bibinfo{author}{Maddulety, K.}, \bibinfo{author}{Bakhshi, P.}, \bibinfo{year}{2024}.
\newblock \bibinfo{title}{Adoption of sustainability in seaport infrastructure: A systematic literature review}.
\newblock \bibinfo{journal}{International Journal of Social Ecology and Sustainable Development (IJSESD)} \bibinfo{volume}{15}, \bibinfo{pages}{1--12}.
\bibitem[{Sharaan et~al.(2024)Sharaan, Ibrahim, Moubarak, ElKut, Romya, Hamouda, Soliman and Iskander}]{sharaan2024qualitative}
\bibinfo{author}{Sharaan, M.}, \bibinfo{author}{Ibrahim, M.G.}, \bibinfo{author}{Moubarak, H.}, \bibinfo{author}{ElKut, A.E.}, \bibinfo{author}{Romya, A.A.}, \bibinfo{author}{Hamouda, M.}, \bibinfo{author}{Soliman, A.}, \bibinfo{author}{Iskander, M.}, \bibinfo{year}{2024}.
\newblock \bibinfo{title}{A qualitative analysis of climate impacts on egyptian ports}.
\newblock \bibinfo{journal}{Sustainability} \bibinfo{volume}{16}, \bibinfo{pages}{1015}.
\bibitem[{Shi et~al.(2023)Shi, Chen, Xu, Di and Qu}]{shi2023improving}
\bibinfo{author}{Shi, J.}, \bibinfo{author}{Chen, J.}, \bibinfo{author}{Xu, L.}, \bibinfo{author}{Di, Z.}, \bibinfo{author}{Qu, Q.}, \bibinfo{year}{2023}.
\newblock \bibinfo{title}{Improving the resilience of maritime supply chains: The integration of ports and inland transporters in duopoly markets}.
\newblock \bibinfo{journal}{Frontiers of Engineering Management} \bibinfo{volume}{10}, \bibinfo{pages}{51--66}.
\bibitem[{Testa et~al.(2015)Testa, Furtado and Alipour}]{testa2015resilience}
\bibinfo{author}{Testa, A.C.}, \bibinfo{author}{Furtado, M.N.}, \bibinfo{author}{Alipour, A.}, \bibinfo{year}{2015}.
\newblock \bibinfo{title}{Resilience of coastal transportation networks faced with extreme climatic events}.
\newblock \bibinfo{journal}{Transportation Research Record} \bibinfo{volume}{2532}, \bibinfo{pages}{29--36}.
\bibitem[{Thakur(2021)}]{thakur2021ports}
\bibinfo{author}{Thakur, S.}, \bibinfo{year}{2021}.
\newblock \bibinfo{title}{Ports and climate uncertainty: An economic imperative for india}.
\newblock \bibinfo{journal}{Maritime Affairs: Journal of the National Maritime Foundation of India} \bibinfo{volume}{17}, \bibinfo{pages}{91--106}.
\bibitem[{Toledo et~al.(2024)Toledo, Pag{\'a}n, L{\'o}pez, Aragon{\'e}s and Olcina}]{toledo2024nature}
\bibinfo{author}{Toledo, I.}, \bibinfo{author}{Pag{\'a}n, J.I.}, \bibinfo{author}{L{\'o}pez, I.}, \bibinfo{author}{Aragon{\'e}s, L.}, \bibinfo{author}{Olcina, J.}, \bibinfo{year}{2024}.
\newblock \bibinfo{title}{Nature-based solutions on the coast in face of climate change: The case of benidorm (spain)}.
\newblock \bibinfo{journal}{Urban Climate} \bibinfo{volume}{53}, \bibinfo{pages}{101816}.
\bibitem[{Trundle et~al.(2019)Trundle, Barth and McEvoy}]{trundle2019leveraging}
\bibinfo{author}{Trundle, A.}, \bibinfo{author}{Barth, B.}, \bibinfo{author}{McEvoy, D.}, \bibinfo{year}{2019}.
\newblock \bibinfo{title}{Leveraging endogenous climate resilience: urban adaptation in pacific small island developing states}.
\newblock \bibinfo{journal}{Environment and Urbanization} \bibinfo{volume}{31}, \bibinfo{pages}{53--74}.
\bibitem[{Tsaimou et~al.(2023)Tsaimou, Sartampakos and Tsoukala}]{tsaimou2023uav}
\bibinfo{author}{Tsaimou, C.N.}, \bibinfo{author}{Sartampakos, P.}, \bibinfo{author}{Tsoukala, V.K.}, \bibinfo{year}{2023}.
\newblock \bibinfo{title}{Uav-driven approach for assisting structural health monitoring of port infrastructure}.
\newblock \bibinfo{journal}{Structure and Infrastructure Engineering} , \bibinfo{pages}{1--20}.
\bibitem[{da~Veiga~Lima and de~Souza(2022)}]{da2022climate}
\bibinfo{author}{da~Veiga~Lima, F.A.}, \bibinfo{author}{de~Souza, D.C.}, \bibinfo{year}{2022}.
\newblock \bibinfo{title}{Climate change, seaports, and coastal management in brazil: An overview of the policy framework}.
\newblock \bibinfo{journal}{Regional Studies in Marine Science} \bibinfo{volume}{52}, \bibinfo{pages}{102365}.
\bibitem[{Verschuur et~al.(2020)Verschuur, Koks and Hall}]{verschuur2020port}
\bibinfo{author}{Verschuur, J.}, \bibinfo{author}{Koks, E.}, \bibinfo{author}{Hall, J.}, \bibinfo{year}{2020}.
\newblock \bibinfo{title}{Port disruptions due to natural disasters: Insights into port and logistics resilience}.
\newblock \bibinfo{journal}{Transportation research part D: transport and environment} \bibinfo{volume}{85}, \bibinfo{pages}{102393}.
\bibitem[{Verschuur et~al.(2022)Verschuur, Pant, Koks and Hall}]{verschuur2022systemic}
\bibinfo{author}{Verschuur, J.}, \bibinfo{author}{Pant, R.}, \bibinfo{author}{Koks, E.}, \bibinfo{author}{Hall, J.}, \bibinfo{year}{2022}.
\newblock \bibinfo{title}{A systemic risk framework to improve the resilience of port and supply-chain networks to natural hazards}.
\newblock \bibinfo{journal}{Maritime Economics \& Logistics} , \bibinfo{pages}{1--18}.
\bibitem[{Vinke et~al.(2022)Vinke, van Koningsveld, van Dorsser, Baart, Van~Gelder and Vellinga}]{vinke2022cascading}
\bibinfo{author}{Vinke, F.}, \bibinfo{author}{van Koningsveld, M.}, \bibinfo{author}{van Dorsser, C.}, \bibinfo{author}{Baart, F.}, \bibinfo{author}{Van~Gelder, P.}, \bibinfo{author}{Vellinga, T.}, \bibinfo{year}{2022}.
\newblock \bibinfo{title}{Cascading effects of sustained low water on inland shipping}.
\newblock \bibinfo{journal}{Climate Risk Management} \bibinfo{volume}{35}, \bibinfo{pages}{100400}.
\bibitem[{Vinke et~al.(2024)Vinke, Turpijn et~al.}]{vinke2024inland}
\bibinfo{author}{Vinke, F.}, \bibinfo{author}{Turpijn, B.}, et~al., \bibinfo{year}{2024}.
\newblock \bibinfo{title}{Inland shipping response to discharge extremes--a 10 years case study of the rhine}.
\newblock \bibinfo{journal}{Climate Risk Management} \bibinfo{volume}{43}, \bibinfo{pages}{100578}.
\bibitem[{Wan et~al.(2022)Wan, Tao, Yang and Zhang}]{wan2022evaluating}
\bibinfo{author}{Wan, C.}, \bibinfo{author}{Tao, J.}, \bibinfo{author}{Yang, Z.}, \bibinfo{author}{Zhang, D.}, \bibinfo{year}{2022}.
\newblock \bibinfo{title}{Evaluating recovery strategies for the disruptions in liner shipping networks: a resilience approach}.
\newblock \bibinfo{journal}{The international journal of logistics management} \bibinfo{volume}{33}, \bibinfo{pages}{389--409}.
\bibitem[{Wan et~al.(2018)Wan, Yang, Zhang, Yan and Fan}]{wan2018resilience}
\bibinfo{author}{Wan, C.}, \bibinfo{author}{Yang, Z.}, \bibinfo{author}{Zhang, D.}, \bibinfo{author}{Yan, X.}, \bibinfo{author}{Fan, S.}, \bibinfo{year}{2018}.
\newblock \bibinfo{title}{Resilience in transportation systems: a systematic review and future directions}.
\newblock \bibinfo{journal}{Transport reviews} \bibinfo{volume}{38}, \bibinfo{pages}{479--498}.
\bibitem[{Wood(2019)}]{wood2019climate}
\bibinfo{author}{Wood, G.D.}, \bibinfo{year}{2019}.
\newblock \bibinfo{title}{Climate delusion: Hurricane sandy, sea level rise, and 1840s catastrophism}.
\newblock \bibinfo{journal}{Humanities} \bibinfo{volume}{8}, \bibinfo{pages}{131}.
\bibitem[{Zhou et~al.(2021)Zhou, Xu, Miller-Hooks, Zhou, Chen, Lee, Chew and Li}]{zhou2021analytics}
\bibinfo{author}{Zhou, C.}, \bibinfo{author}{Xu, J.}, \bibinfo{author}{Miller-Hooks, E.}, \bibinfo{author}{Zhou, W.}, \bibinfo{author}{Chen, C.H.}, \bibinfo{author}{Lee, L.H.}, \bibinfo{author}{Chew, E.P.}, \bibinfo{author}{Li, H.}, \bibinfo{year}{2021}.
\newblock \bibinfo{title}{Analytics with digital-twinning: A decision support system for maintaining a resilient port}.
\newblock \bibinfo{journal}{Decision Support Systems} \bibinfo{volume}{143}, \bibinfo{pages}{113496}.
\bibitem[{Zittis et~al.(2023)Zittis, Ahrens, Obermann-Hellhund, Giannakis, Risto, Agulles~Gamez, Jorda, Quesada~Pe{\~n}a, Lora~Rodr{\'\i}guez, Guersi~Sauret et~al.}]{zittis2023maritime}
\bibinfo{author}{Zittis, G.}, \bibinfo{author}{Ahrens, B.}, \bibinfo{author}{Obermann-Hellhund, A.}, \bibinfo{author}{Giannakis, E.}, \bibinfo{author}{Risto, D.}, \bibinfo{author}{Agulles~Gamez, M.}, \bibinfo{author}{Jorda, G.}, \bibinfo{author}{Quesada~Pe{\~n}a, M.}, \bibinfo{author}{Lora~Rodr{\'\i}guez, V.}, \bibinfo{author}{Guersi~Sauret, J.L.}, et~al., \bibinfo{year}{2023}.
\newblock \bibinfo{title}{Maritime transport and regional climate change impacts in large eu islands and archipelagos}.
\newblock \bibinfo{journal}{Euro-Mediterranean Journal for Environmental Integration} \bibinfo{volume}{8}, \bibinfo{pages}{441--454}.

\end{thebibliography}
\end{document}